\documentclass[english]{article}
\usepackage[T1]{fontenc}
\usepackage[latin9]{inputenc}
\usepackage{geometry}
\usepackage{color}
\usepackage{babel}
\usepackage{float}
\usepackage{amsmath}
\usepackage{amssymb}
\usepackage{graphicx}
\usepackage[pdfusetitle,
 bookmarks=true,bookmarksnumbered=false,bookmarksopen=false,
 breaklinks=false,pdfborder={0 0 0},pdfborderstyle={},backref=false,colorlinks=true]
 {hyperref}

\makeatletter
\newcommand{\lyxaddress}[1]{
	\par {\raggedright #1
	\vspace{1.4em}
	\noindent\par}
}

\numberwithin{equation}{section}

\makeatother

\begin{document}
\title{Comparing the elliptic Ruijsenaars--Schneider model to the finite
volume sine-Gordon theory}
\author{Zoltan Bajnok$^{1,2}$, Apor Roth$^{1,3}$}
\maketitle

\lyxaddress{\begin{center}
\emph{$^{1}$Wigner Research Centre for Physics}\\
\emph{Konkoly-Thege Miklós u. 29-33, 1121 Budapest , Hungary}\\
\emph{$^{2}$Institute of Theoretical Physics and Mark Kac Center
for Complex}\\
\emph{Systems Research, Jagiellonian University, 30-348 Kraków, Poland}\\
$^{3}$\emph{Roland Eötvös University, Department for Theoretical
Physics,}\\
\emph{ 1117 Budapest, Pázmány sétány 1/A, Hungary}
\par\end{center}}
\begin{abstract}
We compare the spectrum of the elliptic Ruijsenaars--Schneider model
with the finite-size spectrum of the sine-Gordon model, highlighting
both their similarities and differences. Our analysis focuses on the
two-particle sector in the center-of-mass frame. At the free point,
we carry out an analytic comparison, while at generic couplings we
employ non-perturbative numerical calculations based on the truncated
Hilbert space method adapted to difference operators. To benchmark
this numerical approach, we first study the trigonometric limit, where
analytic results are available. We then examine in detail the non-relativistic
limit, which encompasses the rational, trigonometric, hyperbolic,
and elliptic Calogero--Moser--Sutherland models. Finally, we compare
the Bethe--Yang momentum quantization conditions, derived from infinite-volume
scattering phases, with the exact finite-volume solutions. We find
that these conditions hold exactly in the rational and trigonometric
cases, but acquire finite-size corrections in the hyperbolic and elliptic
cases, both for the relativistic model and its non-relativistic limit,
however, these corrections are different in quantum field theories
and in many body systems
\end{abstract}

\section{Introduction}

Integrable quantum field theories (QFTs) are important for several
reasons \cite{Dorey:1996gd,Samaj:2013yva,Mussardo2010}. They provide
simplified settings in which strongly coupled, non-perturbative phenomena
can be analysed exactly. While QFTs underpin the fundamental interactions
of nature, they also describe condensed matter systems and play central
roles in statistical physics. In higher dimensions, interacting QFTs
are extremely complicated and generally not exactly solvable. In contrast,
in two space-time dimensions there exist integrable QFTs that, while
still sharing many essential features with realistic systems, remain
analytically tractable. These theories are not only valuable as simplified
laboratories for quantum phenomena but are also of intrinsic mathematical
interest, being deeply connected to representation theory, quantum
groups, and other areas of mathematics.

In integrable theories the scattering matrix is purely elastic: there
is no particle production, and particles retain their velocities during
scattering \cite{Zamolodchikov1979}. This property is already visible
at the classical level. In the sine-Gordon model, the finite-energy
classical solutions consist of solitons, anti-solitons, and their
bound states, the breathers \cite{Rajaraman1982}. During time evolution,
these excitations preserve their velocities. The interaction manifests
only through time delays, which can be obtained by summing the individual
delays a particle would experience when scattering off each of the
others separately. In this integrable field theory there is no dispersion,
and asymptotic solutions consist of well-separated wave fronts that
retain their profiles throughout the scattering process. These localised
excitations can be interpreted as particles.

Ruijsenaars and Schneider posed the question of whether a relativistically
invariant multi-particle system (not a field theory) could reproduce
the sine-Gordon time delays. They answered affirmatively by constructing
the hyperbolic Ruijsenaars--Schneider (RS) model, which is relativistically
invariant and whose time delays coincide exactly with those of the
sine-Gordon model \cite{Ruijsenaars:1986vq}. Later, Ruijsenaars extended
this correspondence to the quantum level, showing that the sine-Gordon
scattering phases can also be reproduced by the quantum RS model \cite{Ruijsenaars:1997aqs},
see also \cite{Hallnas:2017dlr,Belousov:2023kva} and references therein.
This remarkable result suggests that an integrable quantum field theory
can, in certain respects, be equivalent to a quantum mechanical system.
In the present work, we address the question of whether this correspondence
can be extended from infinite to finite volume. The \emph{elliptic
}RS model can be viewed as the finite-volume generalization of the
hyperbolic RS model, (for recent reviews see \cite{Hallnas:2023ozo,Hallnas:2024jhj}).
Our aim is therefore to compare its energy spectrum with that of the
finite-volume sine-Gordon theory.

In quantum field theory, finite-size effects are governed by the scattering
matrix. If the same holds in quantum mechanical systems, then a correspondence
with QFT spectra becomes plausible. In QFT, the leading large-volume
corrections are captured by the Bethe--Yang equations \cite{Yang1969,Zamolodchikov1990}.
These enforce periodicity of the multiparticle wavefunction by incorporating
the scattering phases particles acquire upon crossing one another,
thereby leading to momentum quantization. However, additional corrections
arise from vacuum polarisation: the vacuum is populated by particle--antiparticle
pairs that propagate around the finite spatial volume and annihilate,
producing exponentially suppressed contributions \cite{Luscher1986a,Luscher1986b}.
These effects modify the Bethe--Yang equations; they are known as
Lüscher corrections \cite{Bajnok:2008bm}. When multi-particle processes
are included, the full finite-size spectrum can be obtained through
the thermodynamic Bethe ansatz (TBA) \cite{Zamolodchikov1990}.

For quantum mechanical systems, no analogous analytic framework is
currently available. We therefore rely primarily on numerical comparisons
(with analytic checks at specific points of the parameter space).
To validate our numerical methods, we examine various limiting cases
of the elliptic RS model. These include the non-relativistic Calogero--Moser--Sutherland
models \cite{Calogero:1970nt,sutherland1972exact,sutherland2004beautiful,moser1976three},
as well as the rational, trigonometric, and hyperbolic limits, where
exact results exist for general couplings. For the quantum mechanical
case, we adapt the non-perturbative truncated Hilbert space method,
benchmarking it against exact solutions whenever possible. We then
compare the numerical spectrum of the quantum elliptic RS model with
the spectrum obtained from the nonlinear integral equation (NLIE)
describing the exact finite-volume sine-Gordon theory \cite{Destri:1994bv,Destri:1997yz,Feverati:1998dt}.
Our analysis, restricted to the two-soliton sector, demonstrates that
the two models differ at finite volume.

The paper is organized as follows: In Section 2 we introduce the sine-Gordon
model and describe its finite-size energy spectrum. Section 3 presents
the elliptic RS model, together with its various limits (Calogero--Moser,
rational, trigonometric, and hyperbolic), and recalls the correspondence
between the hyperbolic RS model and the sine-Gordon theory. In Section
4 we introduce the truncated Hilbert space method and apply it to
study the finite-size spectra of the trigonometric and elliptic CMS
systems, analyzing their relation to the Bethe--Yang equations. Section
5 generalizes the method to the relativistic RS cases and Section
6 compares their spectra with those of the finite-volume sine-Gordon
model. Finally, the paper ends with concluding remarks.

\section{Sine-Gordon model}

The sine-Gordon model is an integrable 1+1 dimensional field theory
of a real scalar field with the Lagrangian \cite{Rajaraman1982}
\begin{equation}
{\cal L}=\frac{1}{2}\partial_{\mu}\varphi\partial^{\mu}\varphi-\frac{m^{2}}{\beta^{2}}(1-\cos\beta\varphi)
\end{equation}

\subsection{Classical theory}

The classical theory admits static, finite energy, localised solutions
\begin{equation}
\varphi_{\pm}(x)=\pm\frac{4}{\beta}\arctan\left(e^{m(x-x_{0})}\right)
\end{equation}
 which interpolate between neighbouring minima of the potential. Their
energy density is localised around $x_{0}$ and they are called the
soliton and the anti-soliton. Since the theory is relativistically
invariant, the boost transformation $(x,t)\to(x\cosh\theta+t\sinh\theta,t\cosh\theta+x\sinh\theta)$
leaves the Lagrangian invariant. As a consequence the boosted solutions
$\varphi_{\pm}(x\cosh\theta+t\sinh\theta)$ keep their shape forever
and travel with velocity $v=\tanh\theta$. These solitary wave solutions
of the field theory are called for short solitons. They have localised
energy densities and can be interpreted as particles. The existence
of infinitely many conserved charges implies that one can construct
scattering solutions, which preserve the number of particles and their
velocities. In particular, the soliton-soliton scattering solution
takes the form 
\begin{equation}
\varphi_{++}(x,t)=\frac{4}{\beta}\arctan\left(\frac{e^{x\cosh\theta+t\sinh\theta-x_{1}}+e^{x\cosh\theta-t\sinh\theta-x_{2}}}{1-e^{2x\cosh\theta-x_{1}-x_{2}+2\ln\tanh\theta}}\right)\label{eq:solsol}
\end{equation}
Asymptotically for $t\to\mp\infty$, they describe two solitons in
the center of mass frame moving oppositely with velocities $v=\pm\tanh\theta$.
The difference compared to the free motion is the time delay or space
displacement $\Delta x=2\ln\tanh\theta$ they experienced 
\begin{equation}
\lim_{t\to\pm\infty}\varphi_{++}(x,t)=\varphi_{+}((x\pm\Delta x/2)\cosh\theta+t\sinh\theta)+\varphi_{+}((x\pm\Delta x/2)\cosh\theta-t\sinh\theta)
\end{equation}
The information on the interaction can be encoded into the energy
dependence of this space displacement. Integrability implies that
space displacements of multi-particle states add up and depend only
on the individual two particle interactions. This is a consequence
of the fact that the space displacement depends purely on the rapidity
difference of the particles and not on their actual initial positions
$x_{i}$.

Ruijsenaars and Schneider parametrized the sine-Gordon soliton-soliton
solution as 
\begin{equation}
\varphi_{++}(x,t)=\frac{4}{\beta}\arctan\left(e^{u_{1}(x,t)}\right)+\frac{4}{\beta}\arctan\left(e^{u_{2}(x,t)}\right)\label{eq:solsolu2}
\end{equation}
 and asked the question, what is the effective dynamics of the $u_{i}(x,t)$
``particle'' trajectories. This is how they constructed the hyperbolic
RS models \cite{Ruijsenaars:1986vq}. This theory is a relativistically
invariant multiparticle theory, which reproduces also the multi-particle
time delays of the sine-Gordon theory. Evenmore, its proper quantization
reproduced the exact scattering phases of the quantum sine-Gordon
theory. See section \ref{sec:Ruijsenaars-Schneider-models} for details.

\subsection{Quantum theory}

The quantum sine-Gordon theory is an integrable quantum field theory.
Being such, it is characterized by its particle content and their
scattering matrix \cite{Zamolodchikov1979}. The quantum theory always
contains the soliton and the anti-soliton, while the rest of the spectrum
depends on the coupling constant, which is parametrized as 
\begin{equation}
p=\frac{\beta^{2}}{8\pi-\beta^{2}}
\end{equation}
For $p>1$ the soliton and the anti-soliton repulse each other and
there are no more particles in the spectrum. For $p=1$ these particles
do not interact and we have a free model. For $0<p<1$ the interaction
is attractive: solitons and anti-solitons form neutral boundstates
with masses $m_{n}=2M\sin\frac{\pi pn}{2}$ , where $M$ is the soliton
mass and $n\leq[1/p]$. The soliton and anti-soliton compose into
a mass degenerate doublet, which scatters on itself by a four by four
matrix, which is the function of the rapidity difference. Unitarity,
crossing symmetry and the Yang-Baxter equation fixes the soliton-soliton
scattering to be
\begin{equation}
S_{++}^{++}(\theta)=S_{--}^{--}(\theta)\equiv S(\theta)=-e^{i\chi(\theta)}\quad;\quad\chi(\theta)=\int_{0}^{\infty}dk\frac{\sin k\theta}{k}\frac{\sinh\frac{\pi(p-1)k}{2}}{\sinh\frac{\pi pk}{2}\cosh\frac{\pi k}{2}}
\end{equation}
while the other non-zero scatterings are
\begin{equation}
S_{+-}^{+-}(\theta)=S_{-+}^{-+}(\theta)=S(\theta)\frac{\sinh\frac{\theta}{p}}{\sinh\frac{i\pi-\theta}{p}}\quad;\quad S_{+-}^{-+}(\theta)=S_{-+}^{+-}(\theta)=S(\theta)\frac{\sinh\frac{i\pi}{p}}{\sinh\frac{i\pi-\theta}{p}}
\end{equation}
 The very same scattering coefficients was reproduced by the hyperbolic
quantum Ruijsenaars-Schneider model \cite{Ruijsenaars:1997aqs}. This
established the connection between the two models in infinite volume.
Our aim is to see, how this correspondence could be extended at finite
volume with periodic boundary condition.

\subsection{Quantum theory with perodic boundary condition}

In relativistic quantum field theories the full finite size spectrum
can be described based on the scattering matrix. One distinguished
feature, different from quantum mechanical systems, is the finite
size correction of the vacuum. This is due to the fact that the vacuum
is the sea of virtual particle anti-particle pairs, whos interaction
can be taken into account in the sine-Gordon theory via the Destri
- de Vega equation \cite{Destri:1994bv}
\begin{align}
E_{0}(L) & =-M\int\frac{dx}{2\pi i}\sinh(x+i\eta)\log(1+e^{iZ(x+i\eta)})\nonumber \\
 & \quad+M\int\frac{dx}{2\pi i}\sinh(x-i\eta)\log(1+e^{-iZ(x-i\eta)})
\end{align}
where the counting function $Z(x)$ satisfies 
\begin{align}
Z(x) & =ML\sinh x+\int_{-\infty}^{\infty}\frac{dy}{2\pi i}G(x-y-i\eta)\log(1+e^{iZ(y+i\eta)})\nonumber \\
 & \quad\qquad-\int_{-\infty}^{\infty}\frac{dy}{2\pi i}G(x-y+i\eta)\log(1+e^{-iZ(y-i\eta)})
\end{align}
and $\eta$ is an arbitrary small positive parameter while $M$ is
the soliton mass. The kernel is the derivative of the scattering matrix
phase 
\begin{equation}
G(x)=\partial_{x}\chi(x)=\int_{-\infty}^{\infty}dk\,e^{ikx}\frac{\sinh\frac{\pi(p-1)k}{2}}{2\sinh\frac{\pi pk}{2}\cosh\frac{\pi k}{2}}
\end{equation}

Once we put two solitons with opposite rapidities $\theta$ and $-\theta$
into the vacuum the energy is modified as \cite{Feverati:1998dt}
\begin{align}
E(L) & =2M\cosh\theta-M\int\frac{dx}{2\pi i}\sinh(x+i\eta)\log(1+e^{iZ(x+i\eta)})\nonumber \\
 & \quad+M\int\frac{dx}{2\pi i}\sinh(x-i\eta)\log(1+e^{-iZ(x-i\eta)})
\end{align}
where the counting function $Z(x)$ now describes how the solitons
polarise the interacting vacuum
\begin{align}
Z(x) & =ML\sinh x+\chi(x-\theta)+\int_{-\infty}^{\infty}\frac{dy}{2\pi i}G(x-y-i\eta)\log(1+e^{iZ(y+i\eta)})\nonumber \\
 & \quad\qquad+\chi(x+\theta)-\int_{-\infty}^{\infty}\frac{dy}{2\pi i}G(x-y+i\eta)\log(1+e^{-iZ(y-i\eta)})
\end{align}
the rapidity is determined by the quantisation condition as $Z(\theta)=2\pi(n+\frac{1}{2})$
and $\chi(0)=0$.

These equations can be solved exactly only for $p=1$ when $G(x)=0$.
In general, one can make a large volume expansion. For very large
volumes the integral terms can be neglected and the quantisation condition
\begin{equation}
Z(\theta)=ML\sinh\theta+\chi(2\theta)+\dots=2\pi n+\pi
\end{equation}
is the logarithm of the Bethe--Yang equation, which, for two particles
with opposite rapidities, takes the form 
\begin{equation}
e^{iML\sinh\theta}S_{++}^{++}(2\theta)=1\label{eq:sGBY}
\end{equation}
Its solution provides the quantised rapidities $\theta_{n}$, which
gives the energy as $E(L)=2M\cosh\theta_{n}+O(e^{-ML})$.

One can expand the integral equations systematically to calculate
the exponentially small $e^{-ML}$ corrections. The leading term is
the so called Lüscher correction \cite{Luscher1986a,Luscher1986b}.
Lüscher calculated this correction for a single standing particle,
which was then extended to moving particles in \cite{Klassen:1990ub,Heller:2008at}
and for multiparticle states in \cite{Bajnok:2008bm}. The generic
form of the second order correction is given in \cite{Bombardelli:2013yka}.

In general, the equations cannot be solved analytically, however,
they can be solved very precisely numerically. In so doing, we discretise
the functions on a grid and perform the convolution by Fourier transform,
where the explicit Fourier transform of the kernel can be used.

Technically, we evaluate $iZ(x+i\eta)$ 
\begin{align}
iZ(x+i\eta) & =iML\sinh(x+i\eta)+i\chi(x+i\eta-\theta)+\int_{-\infty}^{\infty}\frac{dx}{2\pi}G(x-y)\log(1+e^{iZ(y+i\eta)})\nonumber \\
 & \quad\qquad+i\chi(x+i\eta+\theta)-\int_{-\infty}^{\infty}\frac{dx}{2\pi}G(x-y+i2\eta)\log(1+e^{-iZ^{*}(y+i\eta)})
\end{align}
where we used that $Z(x-i\eta)=Z^{*}(x+i\eta)$ and obtained a closed
system for $iZ(x+i\eta)$ for a given $\theta$. In evaluating $\chi(\theta)$
and $\chi(x+i\eta)$ we use 
\begin{equation}
i\chi(\theta)=\int_{0}^{\infty}dk\frac{e^{ik\theta}-e^{-ik\theta}}{k}\frac{\sinh\frac{\pi(p-1)k}{2}}{2\sinh\frac{\pi pk}{2}\cosh\frac{\pi k}{2}}
\end{equation}
and 
\begin{equation}
i\chi(x+i\eta)=\int_{-\infty}^{\infty}\frac{dk}{k}\,e^{ikx}\frac{e^{-k\eta}\sinh\frac{\pi(p-1)k}{2}}{2\sinh\frac{\pi pk}{2}\cosh\frac{\pi k}{2}}
\end{equation}
At the quantisation condition we take 
\begin{align}
Z(\theta) & =ML\sinh(\theta)+\int_{-\infty}^{\infty}\frac{dx}{2\pi i}G(\theta-y-i\eta)\log(1+e^{iZ(y+i\eta)})\nonumber \\
 & \quad\qquad+\chi(2\theta)-\int_{-\infty}^{\infty}\frac{dx}{2\pi i}G(\theta-y+i\eta)\log(1+e^{-iZ^{*}(y+i\eta)})
\end{align}
We can solve these equation by iteration with arbitrary precision.
We start from the Bethe--Yang values and update $iZ(x+i\eta)$ and
$\theta$ in each step until we reach the prescribed precision. We
would like to compare the obtained finite volume spectrum to the elliptic
Ruijsenaars-Schneider model, so we recall their definition now.

\section{Ruijsenaars-Schneider models}

\label{sec:Ruijsenaars-Schneider-models}

The RS multiparticle models are very unique of being relativistically
invariant with potential interaction of the same type. Their Hamiltonian
and momentum takes the following form \cite{Ruijsenaars:1986vq,Ruijsenaars:1986pp}
\begin{equation}
H=m\sum_{j=1}^{N}\cosh\theta_{j}\prod_{1\leq j<k\leq N}f(q_{j}-q_{k})\quad;\quad P=m\sum_{j=1}^{N}\sinh\theta_{j}\prod_{1\leq j<k\leq N}f(q_{j}-q_{k})
\end{equation}
where $\theta_{j}$ and $q_{i}$ are canonically conjugated phase
space coordinates: $\{\theta_{j},q_{i}\}=\delta_{ij}$ and we put
the speed of light to $1$. Together with the boost operator $B=-\sum_{i=1}^{N}q_{i}$
they generate the 1+1 dimensional Poincare algebra 
\begin{equation}
\{H,P\}=0\quad;\quad\{H,B\}=P\quad;\quad\{P,B\}=H
\end{equation}
 The first commutator restricts the potential to the form 
\begin{equation}
f(q)^{2}=a+b\wp(q,L,\frac{i\pi}{d})
\end{equation}
where $\wp$ is the Weierstrass $\wp$-function with real periodicity
$L$, and imaginary periodicity $\pi/d$, while $a,b$ are real parameters.
This model is referred to as the elliptic RS model. In the free case,
$f=1$, the form of the energy and momentum indicates that $\theta$
should be interpreted as the rapidity and $q$, being its conjugate,
cannot be the space coordinate. The canonically conjugated momentum
and space coordinates are as $p_{i}=m\sinh\theta_{i}$ and $x_{i}=\frac{q_{i}}{m\cosh\theta_{i}}$
\cite{Ruijsenaars:1986vq}. Thus the coordinate $x_{i}$ seems like
the Lorentz contracted $q_{i}$, implying that the $q_{i}$ phase
space coordinate should be understood as the coordinate in the particle's
own rest frame. In this sense the interaction is rather peculiar depending
on the coordinate difference of the particles' rest coordinates, but
this is the form, which ensures the relativistic invariance. The theory
is actually invariant under a higher symmetry. The quantities 
\begin{equation}
S_{\pm l}=\sum_{\substack{I\subset\{1,\dots,N\}\\
|I|=l
}
}\exp\left(\pm\sum_{j\in I}\theta_{j}\right)\prod_{\substack{j\in I\\
k\notin I
}
}f(q_{j}-q_{k})\,,\label{eq:cons_charge}
\end{equation}
all Poisson commute, including the Hamiltonian and momentum, which
can be expressed as $\frac{m}{2}(S_{1}\pm S_{-1})$, respectively
\cite{Ruijsenaars:1986vq}. This implies that the model is integrable
in the Liouville sense: we have as many conserved charges as many
particles. 

Various limits are interesting. The hyperbolic limit is when the real
period goes to infinity $L\to\infty$: 
\begin{equation}
\wp(q,\infty,\frac{i\pi}{d})=\frac{d^{2}}{3}+\frac{d^{2}}{\sinh^{2}(dq)}\quad;\quad f^{2}(q)=1+\frac{\sin^{2}(d\lambda)}{\sinh^{2}(dq)}
\end{equation}
leading to particles on the whole line. Here we choose specific $a,b$
coefficients. This model is related to the classical sine-Gordon theory
\cite{Ruijsenaars:1986vq}. Let us choose some initial coordinates
on the phase space $u=\{(\theta_{i},q_{i})\}$ and generate a flow
with the time ($H)$ and space ($P)$ evolution generators as $u_{i}(x,t)=(e^{tH-xP}u)\vert_{q_{i}}\equiv q_{i}(x,t)$.
It can be shown that for $d=1/2$ and $\lambda=\pi$
\begin{equation}
\varphi(x,t)=\frac{4}{\beta}\sum_{j=1}^{N}\arctan(e^{u_{j}(x,t)})
\end{equation}
satisfies the sine-Gordon equation of motion. In the two particle
case this leads to the soliton-soliton solution. Observe that the
relation between the particles' physical trajectories and $u_{i}$
is rather involved \cite{Balog:2014rva}. 

In the trigonometric limit the imaginary period goes to infinity $d\to0$,
which formulates the interactions on the circle 
\begin{equation}
\wp(q,L,i\infty)=-\frac{\pi^{2}}{3L^{2}}+\frac{\pi^{2}}{L^{2}\sin^{2}(\frac{\pi}{L}q)}\quad;\quad f^{2}(q)=1+\frac{\sinh^{2}(\pi\lambda/L)}{\sin^{2}(\pi q/L)}
\end{equation}

If we send both periods to infinity we obtain the rational model
\begin{equation}
\wp(x,\infty,i\infty)=\frac{1}{q^{2}}\quad;\quad f^{2}(q)=1+\frac{\lambda^{2}}{q^{2}}
\end{equation}

Let us see now, how we could quantise these non-standard systems.

\subsection{Quantum models}

Following Ruijsenaars, we quantise the canonical variables as $(\theta_{j},q_{j})\to(-i\partial_{q_{j}},q_{j})$.
We set $\hbar=1$ and $m=1$. In the free case $f=1$, the Hamiltonian
takes the form \cite{Ruijsenaars:1986pp}
\begin{equation}
H=\frac{1}{2}\sum_{j=1}^{N}(e^{-i\partial_{q_{j}}}+e^{i\partial_{q_{j}}})\equiv\frac{1}{2}\sum_{j=1}^{N}(T_{j}^{+1}+T_{j}^{-1})
\end{equation}
where the shift operators act as 
\begin{equation}
T_{j}^{\pm1}f(q_{1},\dots,q_{j},\dots,q_{N})=f(q_{1},\dots,q_{j}\mp i,\dots,q_{N})
\end{equation}
The eigenfunctions take the form \cite{Ruijsenaars:1986pp}
\begin{equation}
H\psi(\{q\})=E\psi(\{q\})\quad;\quad\psi(\{q\})=e^{i\sum_{j=1}^{N}\theta_{j}q_{j}}\label{eq:freeRS}
\end{equation}
with the energy and momentum eigenvalues 
\begin{equation}
E=\sum_{i=1}^{N}\cosh\theta_{j}\quad;\quad P=\sum_{i=1}^{N}\cosh\theta_{j}
\end{equation}
Thus the $\theta_{j}$s are the correctly normalised rapidity variables.
The wave function is a bit strange as it contains the $q_{j}$ variables
in the exponents instead of the coordinates $x_{j}$.

In defining the quantum versions of the interacting Hamiltonians we
have to face with ordering problems between $T_{j}^{\pm1}$ and the
potential terms $f(q_{i}-q_{j})$. Ruijsenaars solved this problem
by introducing the factorisation of the potential into $f(q)=f_{+}(q)f_{-}(q)$
and demanding that the appropriately ordered quantum conserved quantities
\begin{equation}
S_{\pm l}=\sum_{\substack{I\subset\{1,\dots,N\}\\
|I|=l
}
}\prod_{\substack{j\in I\\
k\notin I
}
}f_{\mp}(q_{j}-q_{k})T_{j}^{\pm1}\prod_{\substack{j\in I\\
k\notin I
}
}f_{\pm}(q_{j}-q_{k})\,.\label{eq:qcons_charge}
\end{equation}
commute. He found the following expressions \cite{ruijsenaars2001sine,Ruijsenaars:1986pp}:
\begin{equation}
f_{\pm}(q)=\begin{cases}
(1\pm i\frac{\lambda}{q})^{1/2} & \text{rational}\\
\left(\frac{\sinh\left(d(q\pm i\lambda)\right)}{\sinh(dq)}\right)^{1/2} & \text{hyperbolic}\\
\left(\frac{\sin\left(\pi(q\pm i\lambda)/L\right)}{\sin(\pi q/L)}\right)^{1/2} & \text{trigonometric}\\
\left(\frac{\sigma(q\pm i\lambda,L,i\pi/d)}{\sigma(q,L,i\pi/d)}\right)^{1/2} & \text{elliptic}
\end{cases}\label{eq:shiftfunctions}
\end{equation}
where $\sigma(q,\pi/L,id)$ is the quasi-periodic Weierstrass sigma-function.
It is related to the Weierstrsass $\wp$-function by the following
functional relation \cite{whittaker2020course}:
\begin{equation}
\wp(z)-\wp(u)=-\frac{\sigma(z+u)\sigma(z-u)}{\sigma^{2}(z)\sigma^{2}(u)}\,,
\end{equation}
which guarantees $f(q)=f_{+}(q)f_{-}(q)$. The rational, hyperbolic
and trigonometric systems (\ref{eq:shiftfunctions}), can be obtained
from the elliptic one as \cite{Ruijsenaars:1986pp}:
\begin{equation}
\sigma(q,\infty,i\pi/d)=\frac{\sinh dq}{d}\exp\left(-\frac{d^{2}}{6}q^{2}\right)
\end{equation}
\begin{equation}
\sigma(q,L,i\infty)=\frac{\sin\pi q/L}{\pi/L}\exp\left(-\frac{\pi^{2}}{6L^{2}}q^{2}\right)
\end{equation}
\begin{equation}
\sigma(q,\infty,i\infty)=q\,.
\end{equation}

It is instructive to investigate the non-relativistic limit of these
models, the Calogero-Sutherland-Moser systems \cite{sutherland1972exact,sutherland2004beautiful}.
To obtain these systems one can restore the dependence on the speed
of light $q\to mcq$, $\theta\to\theta/mc$, $d\to dmc$, $L\to L/mc$
and take the $c\to\infty$ limit. The result is a usual quantum many
body system, which in the physical coordinates reads as
\begin{equation}
\mathcal{H}=-\frac{1}{2m}\sum_{i=1}^{N}\partial_{x_{i}}^{2}+\frac{\lambda(\lambda-1)}{2m}\sum_{1\leq j<k\leq N}V(x_{i}-x_{j})
\end{equation}
with the potentials \cite{Ruijsenaars:2009}
\begin{equation}
V=\begin{cases}
V_{\mathrm{rat}}=1/x^{2} & \mathrm{rational}\\
V_{\mathrm{hyp}}=d^{2}/\sinh^{2}(dx) & \mathrm{hyperbolic}\\
V_{\mathrm{trig}}=\pi^{2}/\left(L^{2}\sin^{2}(\pi x/L)\right) & \mathrm{trigonometric}\\
V_{\mathrm{ell}}=\wp(x,L,i\pi/d) & \mathrm{elliptic}
\end{cases}\label{eq:CMSpots}
\end{equation}
In the following we recall the solutions of these systems and the
relations between them, together with an efficient numerical approach
to calculate their finite size spectra.

\section{Calogero-Moser-Sutherland systems and the truncated Hilbert space
method}

In this section we investigate the CMS models for two-particles in
the center of mass frame at generic couplings. The appropriately scaled
Hamiltonian leads to the Schrödinger equation 
\begin{equation}
\mathcal{H}\psi\equiv(-\partial_{x}^{2}+\lambda(\lambda-1)V(x))\psi=E\psi
\end{equation}
and describes the motion of a single particle in the same potential.
We are going to analyse sytems in infinite and also in finite volume
with periodic boundary conditions.

In infinite volume all the potentials (\ref{eq:CMSpots}) tend to
zero at infinities implying that the spectrum is continuous and the
wave function behaves asymptotically as in a free theory \cite{vsamaj2008introduction,Samaj:2013yva}:
\begin{equation}
\psi_{x\rightarrow\infty}=\pm\left(e^{-ikx}+S(k)e^{ikx}\right)\quad;\quad\psi_{x\rightarrow-\infty}=e^{-ikx}S(k)+e^{ikx}\,\label{eq:asympgeneral-1}
\end{equation}
Since the potential is symmetric the wave functions can be chosen
to be symmetric or antisymmetric. The appearing scattering phases
$S(k)$ can be interpreted both as transmissions through to the potential
or alternatively as reflections. These are pure phases and carry information
on the interaction. We are going to determine these phases for the
rational and hyperbolic models.

In order to describe the periodic versions of the previous potentials
we formally place infinitely many particles periodically at distance
$L$: $V_{\mathrm{per}}(x)=\sum_{n=-\infty}^{\infty}V(x+nL)$. We
demand the wave functions to be periodic as well $\psi(x+L)=\psi(x)\,.$This
means that we analyse the periodic system and not the lattice of particles,
which comes with quasi periodic boundary condition and would lead
to finite band solutions. Contrary, periodicity implies momentum quantisation
and leads to discrete energy spectrum.

\begin{figure}
\centering{}\includegraphics[width=8cm]{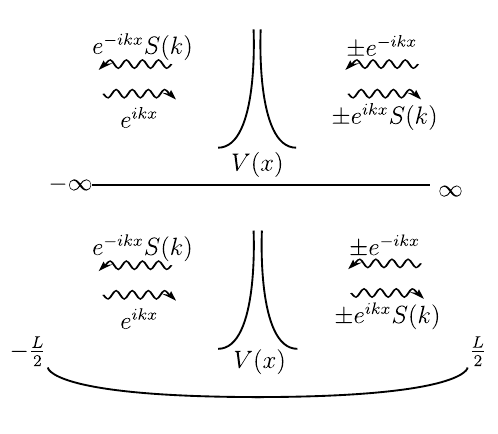}\caption{Justification of the Bethe--Yang equations for odd and even wavefunctions.
The asymptotic behaviour of the wave function is indicated in infinite
volume above. For a large volume system we use the same asymptotic
behaviour, and demand the periodicity of the wavefunction.}
\label{fig:ba}
\end{figure}

A large volume approximation of the spectrum can be obtained from
the scattering phase via the Bethe-Yang equation 
\begin{equation}
e^{ikL}S(k)=\pm1\quad;\quad E=k^{2}
\end{equation}
where the different signs appear for the symmetric and anti-symmetric
cases. This picture assumes that we have only one copy of the potential,
which is manifested in the scattering phase and periodicity is taken
into account for the asymptotic behaviour. Clearly this is only an
approximation, which is demonstrated on Figure \ref{fig:ba}. Typically,
there are finite size corrections to this BY equations, which in the
relativistic QFT case, can be calculated from the logarithmic derivative
of the scattering phase \cite{Zamolodchikov1990}.

The finite size spectrum can also be calculated numerically using
the truncated Hilbert space method. This is a non-perturbative variational
approach, which is based on splitting the Hamiltonian into a soluble
and a perturbation part 
\begin{equation}
\mathcal{H}=\mathcal{H}_{0}+\mathcal{H}_{\mathrm{pert}}=-\frac{\partial^{2}}{\partial x^{2}}+V_{\mathrm{per}}(x),
\end{equation}
and using the eigensystem of the free part as the variational basis.
Since the spectrum is discrete it can be easily truncated at a given
free eigenenergy and the variational method boils down to diagonalizing
the truncated Hamiltonian. The free part has even\footnote{The $n=0$ case should be normalised as $\Phi_{0}^{(+)}=L^{-1/2}$.}
and odd eigenfunctions with the energies 
\begin{equation}
\Phi_{n}^{(+)}=\sqrt{\frac{2}{L}}\cos\left(\frac{2\pi nx}{L}\right)\quad;\quad\Phi_{n}^{(-)}=\sqrt{\frac{2}{L}}\sin\frac{2\pi nx}{L}\quad;\quad\mathcal{H}_{0}\Phi_{n}^{(\pm)}=\left(\frac{2\pi n}{L}\right)^{2}\Phi_{n}^{(\pm)}.\label{eq:simplebasis}
\end{equation}
We then calculate the matrix elements of the perturbation $V_{nm}^{(\pm)}=\langle\Phi_{n}^{(\pm)}|\hat{\mathcal{H}}_{\mathrm{pert}}|\Phi_{m}^{(\pm)}\rangle$,
where due to the symmetry of the potential the even and odd sector
is orthogonal to each other. By truncating the Hilbert space with
$n\leq N$ we have to diagonalise a finite matrix eventually
\begin{equation}
H_{nm}=\left(\frac{2\pi n}{L}\right)^{2}\delta_{nm}+V_{nm}\,.
\end{equation}
If we increase the truncation level $N$ we get a better and better
spectrum.

\subsection{The rational CMS system}

\begin{figure}[H]
\begin{centering}
\includegraphics[width=14cm]{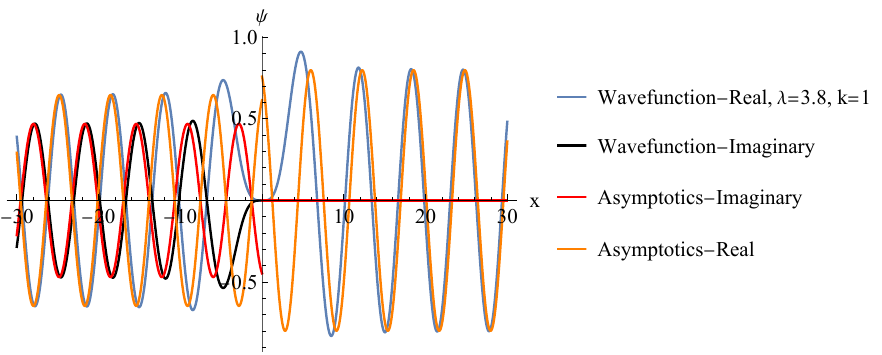}
\par\end{centering}
\caption{Solution for the rational potential for $\lambda=3.8$ and $k=1$.
One can see, that the wavefunction approaches to the asymptotic wavefunctions
for large positive and negative values.}

\label{fig:bessel}
\end{figure}

This model describes particles on the whole line, which interact via
the potential $V_{\mathrm{rat}}$. The Schrödinger equation in the
center of mass frame 
\begin{equation}
-\frac{\partial^{2}\psi(x)}{\partial x^{2}}+\frac{\lambda(\lambda-1)}{x^{2}}\psi(x)=k^{2}\psi(x)\,.
\end{equation}
 has two solutions out of which 
\begin{equation}
\psi(x)=\sqrt{x}J_{\lambda-1/2}(kx)
\end{equation}
is $\delta-$normalisable \cite{vsamaj2008introduction,Samaj:2013yva}.
By inspecting the asymptotic behaviour of the Bessel functions, the
scattering phase can be extracted 
\begin{equation}
S(k)=e^{-i\pi\lambda\,\mathrm{sgn}(k)}
\end{equation}
It is instructive to compare the asymptotic solution with the exact
one, which is done on Figure \ref{fig:bessel}. Clearly, the wavefunctions
agree only asymptotically. Note also that the physical solution is
zero at the origin, which is required by its square integrability
around a small region including zero in the potential with $1/x^{2}$
singularity.

\subsection{The trigonometric CMS system}

Let us now examine the periodic version of the inverse square potential.
This interaction potential can be obtained by summing up the contributions
of periodically placed particles of disctance $L$: $\sum_{n=-\infty}^{\infty}\frac{1}{(x+nL)^{2}}=\left(\frac{\pi}{L}\right)^{2}\frac{1}{\sin^{2}(\frac{\pi x}{L})}$
and leads to the following Schrödinger equation 
\begin{equation}
-\frac{\partial^{2}\psi(x)}{\partial x^{2}}+\left(\frac{\pi}{L}\right)^{2}\frac{\lambda(\lambda-1)}{\sin^{2}(\frac{\pi x}{L})}\psi(x)=k^{2}\psi(x)\,.\label{eq:nr_sin}
\end{equation}
This equation also defines the interaction of particles on the circle.
In order to find physical solutions we need to specify boundary conditions.
If we are interested in the motion of a particle in the periodically
placed particles potential we need to demand quasi periodicity, which
leads to Bloch waves and band solutions, similarly to a solid state
system. In contrast, here we are interested in the $1/\sin^{2}$ interaction
on the circle thus we demand the periodicity for the wave function.
This system allows an analytical solution, which we present now for
the symmetric and anti-symmetric sector. In the symmetric case first
observe, that the groundstate solution is simply \cite{sutherland1972exact}
\begin{equation}
\psi_{0}^{(+)}=\left|\sin\left(\frac{\pi}{L}x\right)\right|^{\lambda}\quad;\quad E_{0}^{(+)}=k_{0}^{2}=\left(\frac{\pi\lambda}{L}\right)^{2}
\end{equation}
We can then look for other symmetric solutions in the form $\psi=\psi_{0}^{(+)}\tilde{\psi}$
and formulate the eigenvalue problem for $\tilde{E}=E-E_{0}$ \cite{sutherland1972exact}:
\begin{equation}
\mathcal{H}^{(+)}\tilde{\psi}=-\frac{\partial^{2}\tilde{\psi}}{\partial x^{2}}-2\frac{\pi}{L}\lambda\cot\left(\frac{\pi}{L}x\right)\frac{\partial\tilde{\psi}}{\partial x}=\tilde{E}\tilde{\psi}\,.
\end{equation}
The reduced Hamiltonian ${\cal H}^{(+)}$ acts in an upper triangular
way in the symmetric basis: 
\begin{equation}
\mathcal{H}^{(+)}\Phi_{n}^{(+)}=\left(\frac{2\pi}{L}\right)^{2}\left(n^{2}\Phi_{n}^{(+)}+n\lambda(\Phi_{n}^{(+)}+2\Phi_{n-1}^{(+)}+2\Phi_{n-2}^{(+)}+\dots+2\Phi_{1}^{(+)}+\Phi_{0}^{(+)})\right)\,,
\end{equation}
The eigenvectors can be calculated iteratively and the eigenvalues
of the full Hamiltonian are 
\begin{equation}
E_{n}^{(+)}=E_{0}^{(+)}+\tilde{E}=k_{n}^{2}=\left(\frac{\pi}{L}\right)^{2}(2n+\lambda)^{2}
\end{equation}

In the antisymmetric sector we construct the odd groundstate 
\begin{equation}
\psi_{0}^{(-)}=\mathrm{sgn}(x)\cos\left(\frac{\pi x}{L}\right)\left|\sin\left(\frac{\pi x}{L}\right)\right|^{\lambda}\quad;\quad E_{0}^{(-)}=\left(\frac{\pi(\lambda+1)}{L}\right)^{2}\,
\end{equation}
and look for the eigenfunctions in the form $\psi=\psi_{0}^{(-)}\tilde{\psi}$
with $\tilde{E}=E-E_{0}^{(-)}$. Although $\mathrm{sgn}(x)\cos\left(\frac{\pi x}{L}\right)$
has a finite jump at the origin, the zero of $\left|\sin\left(\frac{\pi x}{L}\right)\right|^{\lambda}$
makes it continues and vanishing there. Similarly to the even case
the reduced Hamiltonian 
\begin{equation}
\mathcal{H}^{(-)}\tilde{\psi}=-\frac{\partial^{2}\tilde{\psi}}{\partial x^{2}}-2\frac{\pi}{L}\left(\lambda\cot\left(\frac{\pi}{L}x\right)-\tan\left(\frac{\pi}{L}x\right)\right)\frac{\partial\tilde{\psi}}{\partial x}=\tilde{E}\tilde{\psi}\,.
\end{equation}
acts an upper triangular way on the even basis, leading to the odd
eigenvalues
\begin{equation}
E_{n}^{(-)}=E_{0}^{(-)}+\tilde{E}=\left(\frac{\pi}{L}\right)^{2}(2n+1+\lambda)^{2}
\end{equation}
which nicely complement the even ones. The eigenvalues corresponding
to odd states can also be obtained in the two-particle formalism in
\cite{vsamaj2008introduction,sutherland1972exact}.

\subsubsection{Finite size corrections and Bethe-Yang equation}

We can now compare the exact finite volume wave numbers $k=(\pi/L)(n+\lambda)$
to the one coming from the BY equation 
\begin{equation}
e^{ikL}e^{-i\pi\lambda}=\pm1
\end{equation}
Clearly the even sector is completely reproduced from the symmetric,
while the odd sector from the anti-symmetric BY equations. This is
suprising as the wavefunctions are really different. In a QFT the
finite size corrections beyond the BY equations are described by the
logarithmic derivative of the scattering phase. This quantity is zero
here, which indicates that corrections in the non-relativistic theories
might have the same origin.

\subsubsection{Truncated Hilbert space approach for the trigonometric CMS model}

Let us solve eq. (\ref{eq:nr_sin}) numerically. In using the truncated
Hilbert space method we split the Hamiltonian as 
\begin{equation}
\mathcal{H}=\mathcal{H}_{0}+\mathcal{H}_{\mathrm{pert}}=-\frac{\partial^{2}}{\partial x^{2}}+\left(\frac{\pi}{L}\right)^{2}\frac{\lambda(\lambda-1)}{\sin^{2}(\pi x/L)}
\end{equation}
The matrix elements in the odd sector are
\begin{equation}
V_{nm}^{(-)}=\left(\frac{\pi}{L}\right)^{2}\lambda(\lambda-1)\frac{2}{L}\int_{-L/2}^{L/2}\frac{\sin\left(\frac{2\pi n}{L}x\right)\sin\left(\frac{2\pi m}{L}x\right)}{\sin^{2}\left(\frac{\pi}{L}x\right)}=4\left(\frac{\pi}{L}\right)^{2}\lambda(\lambda-1)\min(n,m)\,.
\end{equation}
In the even sector some matrix elements are divergent, so we regularise
them as 
\begin{equation}
V_{nm}^{(+)}=2\int_{\varepsilon}^{L/2}\Phi_{n}^{(+)}{\cal H}_{\mathrm{pert}}\Phi_{m}^{(+)}dx\,.
\end{equation}
This way only one eigenvalue got corrupted with a very high energy
and we could calculate the first $N$ eigenvalues and eigenfunctions.
Actually, the singular subspace can be easily identified and one can
choose a basis orthogonal to this subspace as 
\begin{equation}
\Phi_{n}=\sqrt{\frac{2}{L}}\sin^{2}\left(\frac{n\pi x}{L}\right)\,.\label{eq:basis_new}
\end{equation}
We can calculate the unperturbed energies and the matrix elements
of the potential for the new basis:
\begin{equation}
H_{0,nm}=\langle\Phi_{n}|{\cal H}_{0}|\Phi_{m}\rangle=\left(\frac{\pi n}{L}\right)^{2}\delta_{nm}
\end{equation}
and 
\begin{equation}
V_{nm}=\left(\frac{\pi}{L}\right)^{2}\lambda(\lambda-1)\frac{2}{L}\int_{-L/2}^{L/2}\frac{\sin\left(\frac{\pi n}{L}x\right)^{2}\sin\left(\frac{\pi m}{L}x\right)^{2}}{\sin^{2}\left(\frac{\pi}{L}x\right)}=\left(\frac{\pi}{L}\right)^{2}\lambda(\lambda-1)\min(n,m)\,.
\end{equation}
As the basis elements are not orthonormal, we need to calculate their
inner product matrix and its inverse
\begin{equation}
\Phi_{kl}=\langle\Phi_{k}|\Phi_{l}\rangle=\frac{1}{4}\delta_{kl}+\frac{1}{2}\quad;\quad\left(\Phi_{kl}\right)^{-1}=4\delta_{kl}-\frac{8}{2N+1}\,.\label{eq:ip}
\end{equation}
In order to obtain the spectrum we need to diagonalise the following
matrix:
\begin{equation}
H_{nm}=\left(\Phi_{nm}\right)^{-1}\cdot\left(H_{0,nm}+V_{nm}\right)\,.
\end{equation}

\begin{figure}[H]
\centering{}\includegraphics[scale=0.75]{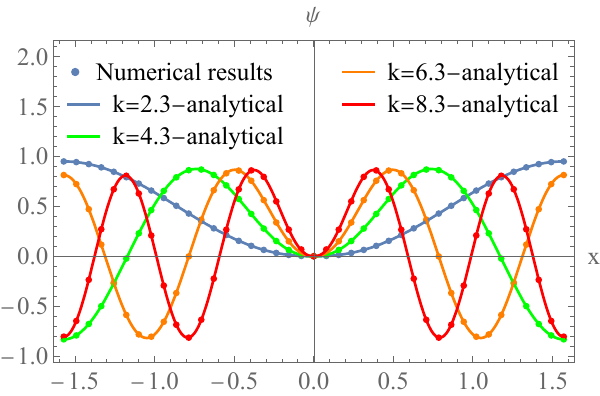}\includegraphics[scale=0.75]{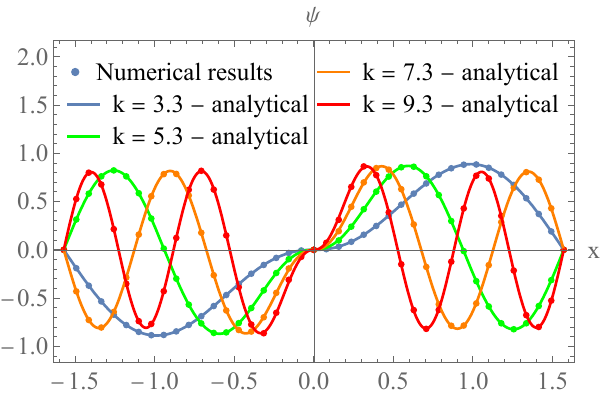}\caption{The first four even (on the left) and odd (on the right) eigenstates
with their eigenvalues calculated analytically and numerically for
$L=\pi$, $\lambda=2.3$, $N=10$.}
\label{sina}
\end{figure}

After calculating the eigenvalues and eigenstates numerically we obtain
a good agreement between the analytical and numerical values. We plotted
the first four eigenstates ( analytical and numerical) for $\lambda=2.3$
in figure \ref{sina}. 

\begin{figure}[h]
\begin{centering}
\includegraphics[width=7cm]{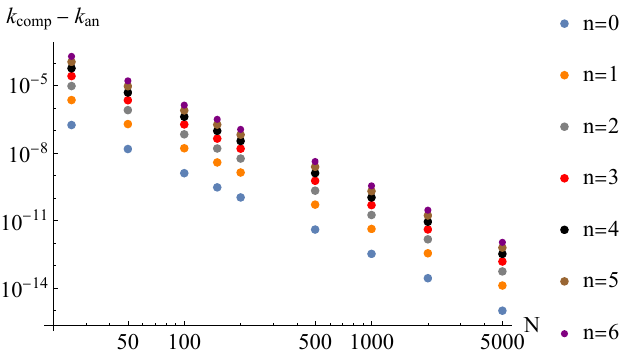}\hspace{1cm}\includegraphics[width=7cm]{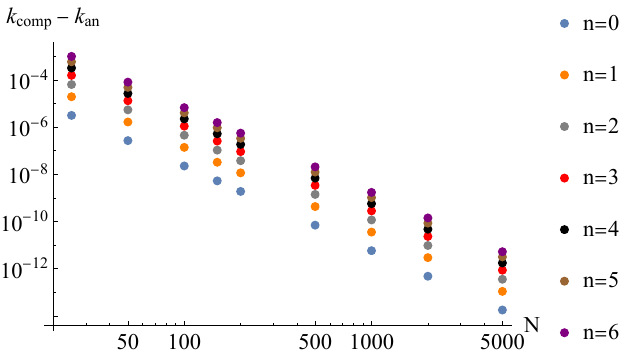}
\par\end{centering}
\caption{Difference between analytical and numerical results for the first
seven even (on the left) and odd (on the right) eigenvalues for different
number of basis elements ($N$) and $\lambda=2.3$. During our numerical
calculations, we calculated the ten lowest eigenvalues using the Arnoldi
method, and we compared the first seven eigenvalues to the analytical
values.}

\label{fig:evenerrsin}
\end{figure}

Next, we should check the convergence of our numerical method. To
do this, we keep $\lambda=2.3$ and increase the number of basis elements
from $N=25$. We examined the convergence for the even and odd bases:
we calculated the first seven wavenumbers numerically for different
number of basis elements and compared to the analytical values. In
figures \ref{fig:evenerrsin} we can see, that the computed wavenumbers
nicely converge to the analytically calculated ones.

\subsection{The hyperbolic CMS system}

\begin{figure}[h]
\begin{centering}
\includegraphics[width=14cm]{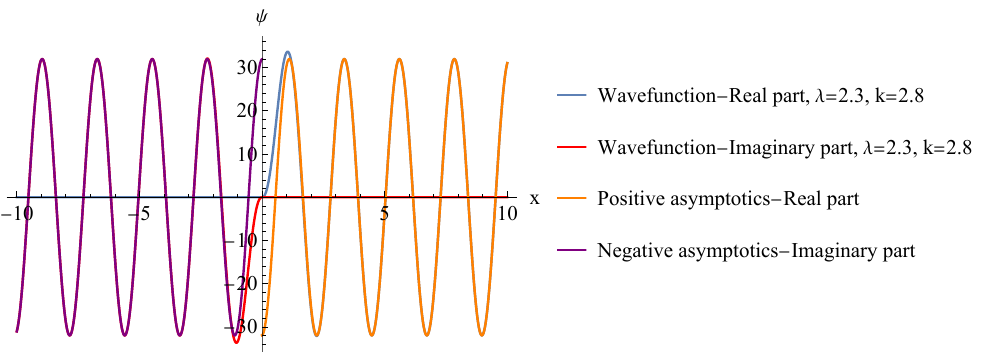}
\par\end{centering}
\caption{Solutions for the hyperbolic potential for $\lambda=2.3$ and $k=2.8$}
\label{fig:hyp}
\end{figure}

The hyperbolic CMS system can be obtained as an analytical continuation
of the trigonometric one with the subtitution of $\frac{\pi}{L}\rightarrow id$
\cite{sutherland2004beautiful,vsamaj2008introduction}. We get the
following Schrödinger equation:
\begin{equation}
-\frac{\partial^{2}\psi(x)}{\partial x^{2}}+d^{2}\frac{\lambda(\lambda-1)}{\sinh^{2}(dx)}\psi(x)=E\psi(x)\,.
\end{equation}
Note that the potential is now periodic in the imaginary direction
with a period of $\frac{\pi}{d}$. In terms of the momentum $k=\sqrt{E}$
we obtain the following $\delta$-normalisable solution \cite{sutherland2004beautiful}:
\begin{equation}
\psi=\,_{2}F_{1}\left(\frac{1}{2}(\lambda-ik/d),\frac{1}{2}(1+\lambda-ik/d),\frac{1}{2}+\lambda,y{}^{2}\right)\frac{1}{\sqrt{y}}(y^{2}-1)^{-\frac{ik}{2d}}(y^{2}){}^{\frac{1}{4}+\frac{\lambda}{2}}
\end{equation}
Here, $y=\tanh dx$ and $_{2}F_{1}$ is the hypergeometric function.
We can multiply the solution in the negative range with a constant
phase factor to get even or odd solutions. We now investigate the
asymptotics and extract the scattering phase. For $x\to\infty$ we
obtain 
\begin{equation}
\psi_{x\rightarrow\infty}\approx\frac{2^{-1+\lambda}\Gamma(\frac{1}{2}+\lambda)e^{\frac{k\pi}{2d}}}{\sqrt{\pi}\Gamma(\lambda-ik/d)\Gamma(\lambda+ik/d)}\left(\Gamma\left(\lambda+ik/d\right)\Gamma\left(-ik/d\right)e^{-ikx}+\Gamma\left(\lambda-ik/d\right)\Gamma\left(ik/d\right)e^{ikx}\right)
\end{equation}
For $x\to-\infty$, up to a constant phase we got the $x\to-x$ transformed
solution, which together give the scattering phase \cite{sutherland2004beautiful,arutyunov2019elements}
\begin{equation}
S(k)=\frac{\Gamma(\lambda-ik/d)\Gamma(ik/d)}{\Gamma(\lambda+ik/d)\Gamma(-ik/d)}\,.\label{eq:bahyp}
\end{equation}
The solution and its asymptotics for $\lambda=2.3$ and $k=2.8$ are
shown in figure \ref{fig:hyp}.

\subsection{The elliptic CMS model}

The elliptic CMS model can be thought of as the periodised version
of the hyperbolic model 
\begin{equation}
\sum_{n=-\infty}^{\infty}\frac{d^{2}}{\sinh^{2}(d(x+Ln))}=\wp(x,L,\frac{i\pi}{d})+\frac{2d\eta_{3}}{i\pi}
\end{equation}
The constant $\eta_{3}$ is the imaginary half-period value of the
quasi-periodic Weierstrass zeta function, i.e. $\eta_{3}=\zeta(\frac{i\pi}{2d})$.
For numerical purposes we can also express the potential in terms
of the $q$-polygamma function $\psi_{q}^{(n)}$, i.e. the $n$-th
derivative of the $q$-digamma funcion: 
\begin{equation}
\sum_{n=-\infty}^{\infty}\frac{d^{2}}{\sinh^{2}\left(d(x+Ln)\right)}=\frac{-4dL+\psi_{e^{dL}}^{(1)}(\frac{x}{L})+\psi_{e^{dL}}^{(1)}(\frac{x}{L}-\frac{i\pi}{dL})+\psi_{e^{dL}}^{(1)}(1-\frac{x}{L})+\psi_{e^{dL}}^{(1)}(1-\frac{x}{L}-\frac{i\pi}{dL})}{L^{2}}
\end{equation}

This model was solved for generic coupling by Langmann in \cite{langmann2014explicit},
perturbatively. For integer $\lambda$-s it is a form of the Lamé
equation \cite{whittaker2020course} and the solution can be written
(for $\lambda=2$) in a compact form (\ref{eq:lambda2ellCMS}). However,
we present here a non-perturbative approach for general coupling.

We need to solve the following equation:
\begin{equation}
-\frac{\partial^{2}\psi(x)}{\partial x^{2}}+\left(\wp(x,L,\frac{i\pi}{d})+\frac{2d\eta_{3}}{i\pi}\right)\lambda(\lambda-1)\psi(x)=E\psi(x)\,,\label{eq:HellCMS}
\end{equation}
and we impose periodic boundary condition $\psi(x+L)=\psi(x)$. First,
we express the Weierstrass function with the help of the Jacobi elliptic
sine function 
\begin{equation}
\wp(x)=e_{3}+\frac{e_{1}-e_{3}}{\text{sn}^{2}(\sqrt{e_{1}-e_{3}}x,m)}\,.\label{eq:weierstrtojacobi}
\end{equation}
where the Weierstrass function values at half periods are $e_{1}=\wp(\frac{L}{2})$,
$e_{2}=\wp(\frac{i\pi}{2d})$ and $e_{3}=\wp(\frac{L}{2}+\frac{i\pi}{2d})$
and $\text{sn}(\sqrt{e_{1}-e_{3}}x,m)$ is the elliptic sine function
with the module square $m=\frac{e_{2}-e_{3}}{e_{1}-e_{3}}$. After
changing variable to $z=\sqrt{e_{1}-e_{3}}x$ we obtain the following
equation
\begin{equation}
-\frac{\partial^{2}\psi(z)}{\partial z^{2}}+\frac{\lambda(\lambda-1)}{\text{sn}^{2}(z,m)}\psi(z)=\left(\frac{E}{e_{1}-e_{3}}-\frac{\lambda(\lambda-1)e_{3}}{e_{1}-e_{3}}-\frac{\lambda(\lambda-1)2d\eta_{3}}{i\pi(e_{1}-e_{3})}\right)\psi(z)\equiv h\psi(z)\,.\label{eq:sninverse}
\end{equation}
where the periodicity requirement takes the form $\psi(z+\sqrt{e_{1}-e_{3}}L)=\psi(z)$.
This further can be transformed into a Heun equation. In doing so,
we write the wavefunction as
\begin{equation}
\psi(z)=|\mathrm{sn}(z)|^{\lambda}\tilde{\psi}(z)\,.
\end{equation}
where here and from now on we do not write out explicitly the parameter
of the Jacobi elliptic function, whenever it leads to no confusion.
We then get an equation for $\tilde{\psi}(z)$: 
\begin{equation}
-\frac{\partial^{2}\tilde{\psi}(z)}{\partial z^{2}}-2\lambda\frac{\text{cn}(z)\:\text{dn}(z)}{\text{sn}(z)}\frac{\partial\tilde{\psi}(z)}{\partial z}+\lambda^{2}(m+1)\tilde{\psi}(z)-\lambda(\lambda+1)m\,\text{sn}^{2}(z))\tilde{\psi}(z)=h\tilde{\psi}(z)\,.
\end{equation}
Here we used that $\text{cn}^{2}z=1-\text{sn}^{2}z$ and $\text{dn}^{2}z=1-m\,\text{sn}^{2}z$
\cite{whittaker2020course}. Finally, by substituting $u=\mathrm{sn}^{2}(z)$
we get the form the Heun equation \cite{ronveaux1995heun}:
\begin{equation}
\frac{\partial^{2}\tilde{\psi}(u)}{\partial u^{2}}+\left(\frac{\gamma}{u}+\frac{\delta}{u-1}+\frac{\epsilon}{u-a}\right)\frac{\partial\tilde{\psi}(u)}{\partial u}+\frac{\alpha\beta u-q}{u(u-1)(u-a)}\tilde{\psi}(u)=0\,
\end{equation}
with the constraint \cite{ronveaux1995heun}: $\epsilon=\alpha+\beta-\gamma-\delta+1$,
where $a=\frac{1}{m}$, $q=\frac{\lambda^{2}(1+\frac{1}{m})-\frac{h}{m}}{4}$,
$\alpha=\frac{\lambda}{2}$, $\beta=\frac{\lambda+1}{2}$, $\gamma=\frac{1+2\lambda}{2}$,
$\delta=\frac{1}{2}$ and $\epsilon=\frac{1}{2}$. We can check that
the parameters satisfy the constraint relation. The solution of the
spectral problem for generic coupling then have the form
\begin{equation}
\psi(z)=|\mathrm{sn}(z)|^{\lambda}Hl\left(\frac{1}{m},\frac{\lambda^{2}(1+\frac{1}{m})-\frac{h}{m}}{4},\frac{\lambda}{2},\frac{\lambda+1}{2},\frac{1+2\lambda}{2},\frac{1}{2},\text{sn}^{2}(z)\right)\,.\label{eq:heunevensmall}
\end{equation}
Written in terms of ${\rm sn}(z)$ the solution is automatically periodic,
but not necessarily regular. Indeed, the quantisation of the momentum
comes from requiring regularity. We elaborate on this in the Appendix.
For integer $\lambda$ we also connect the eigenvalues to the eigenvalues
of the Lame equation. 
\begin{equation}
-\frac{\partial^{2}y(\tilde{z})}{\partial\tilde{z}^{2}}+\lambda(\lambda-1)m\:\text{sn}^{2}(\tilde{z})y(\tilde{z})=hy(\tilde{z})\,.
\end{equation}
where $\tilde{z}=z+i\frac{\sqrt{e_{1}-e_{3}}\pi}{d}$ . We have to
distinguish between even and odd $\lambda$s. The eigenvalues of the
even problem are denoted by $b_{\lambda-1}^{2s+2}$, while the corresponding
eigenfunctions as $Es_{\lambda-1}^{2s+2}(\tilde{z})$ \cite{ince1940v},
while in the odd $\lambda$ case the eigenvalues are denoted as $a_{\lambda-1}^{2s+1}$
and correspond to the Lamé functions $Ec_{\lambda-1}^{2s+1}(\tilde{z})$
\cite{ince1940v}. Here $s$ labels the solutions.

For integer $\lambda$s the Lame solutions can also be written in
terms of Weierstrass $\sigma$- and $\zeta$-functions . For comparison
we present here the simplest solution for (\ref{eq:HellCMS}), which
appears for $\lambda=2$ \cite{whittaker1927,RuijGenLame,Klabbers:2020osb}.
\begin{equation}
\psi(x)=\frac{\sigma(x+z)}{\sigma(x)}e^{-\zeta(z)x}\label{eq:lambda2ellCMS}
\end{equation}
and has energy $E=-\wp(z)+\frac{4d\eta_{3}}{i\pi}$. We can ensure
periodicity $\psi(x)=\psi(x+L)$ by looking for $z$ in the form $z=i\alpha_{n}$,
and demanding 
\begin{equation}
\eta_{1}\alpha_{n}+i\zeta(i\alpha_{n})\omega_{1}=\pi n
\end{equation}
which has a solution for any $n$. The physical solution is $\psi(x)+\psi(-x)$,
which we compared against the Lame and Heun solutions for this particular
$\lambda$. The corresponding energy is $E_{n}=-\wp(i\alpha_{n})+\frac{4d\eta_{3}}{i\pi}$.

\subsection{Numerical examination of the elliptic Calogero-Moser system}

\begin{figure}[H]
\centering{}\includegraphics[width=7cm]{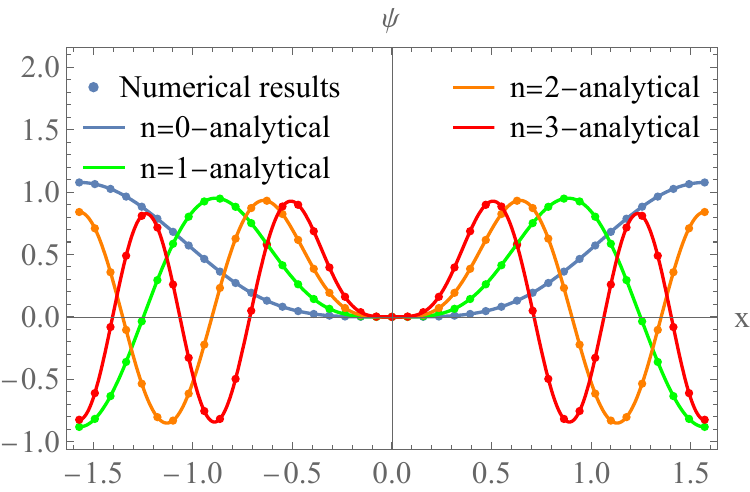}\hspace{1cm}\includegraphics[width=7cm]{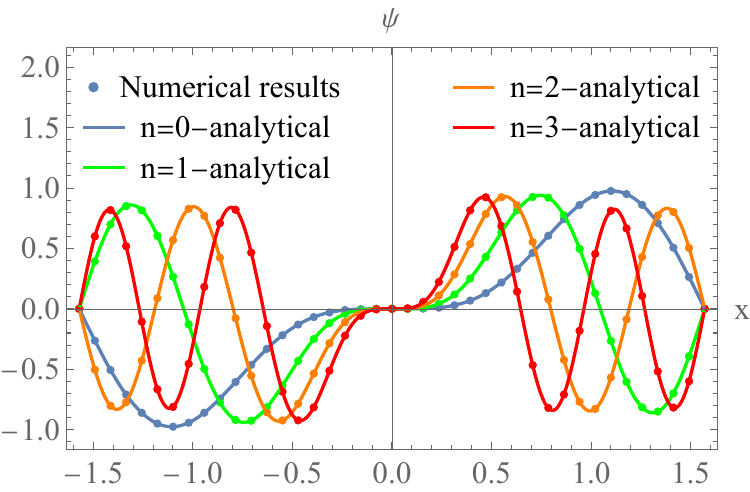}\caption{The first four even eigenstates with their eigenvalues calculated
analytically and numerically and normalised for $L=\pi$, $d=1$,
$\lambda=4$, $N=10$.}
\label{fig:numhyp}
\end{figure}

Let us move to the numerical analysis of the system. We use the method
we highlighted at the beginning of the section. Although the method
works for any $\lambda$, we present the results for integer $\lambda$s,
where comparison with the Lame solution is also available. For $\lambda=4$,
$L=\pi$ and $d=1$ we calculated the first four even and odd eigenstates
and compared them to eq. (\ref{eq:heunevensmall}) and eq. (\ref{eq:heunoddsmall-1})
for even and odd eigenstates in figure \ref{fig:numhyp}.

We also tested the convergence of our numerical solutions by calculating
the first seven eigenvalues for different number of basis elements,
and we could see that the difference between the momenta, calculated
from Lamé eigenvalues, and the momenta, calculated numerically, decreases
systematically when increasing the number of basis elements.

\begin{figure}[H]
\begin{centering}
\includegraphics[width=7cm]{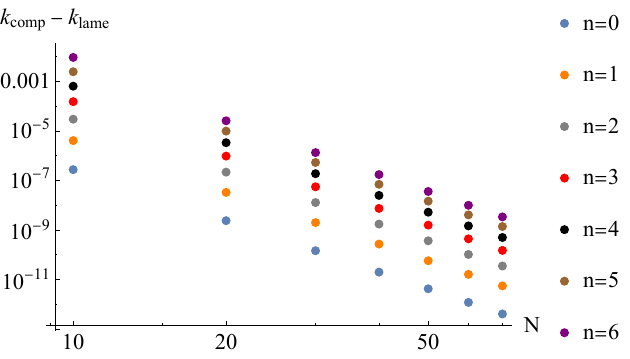}\hspace{1cm}\includegraphics[width=7cm]{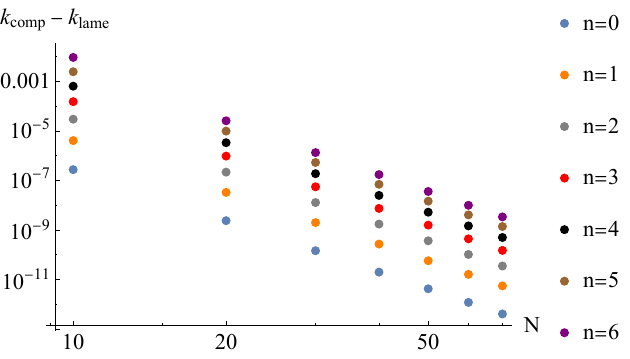}
\par\end{centering}
\caption{Difference between analytical and numerical results for the first
seven even ( $\lambda=3$, on the left) and odd ( $\lambda=4$, on
the right) momenta for different number of basis elements ($N$) for
the hyperbolic potential.}

\label{fig:evenerrshyp}
\end{figure}

\subsection{Comparison between the Bethe--Yang equation and the wavenumbers
of the elliptic potential}

\begin{figure}[H]
\begin{centering}
\includegraphics[width=7cm]{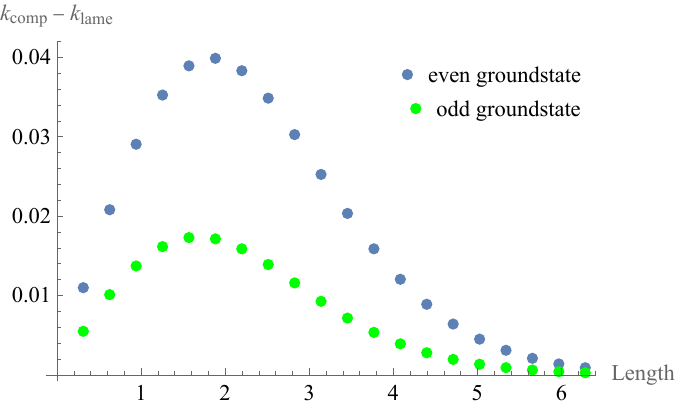}\hspace{1cm}\includegraphics[width=7cm]{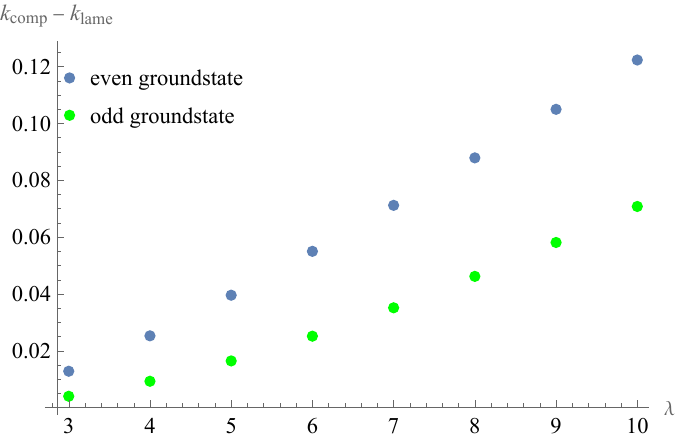}
\par\end{centering}
\caption{Validity of the Bethe--Yang equations for the even and odd ground
states at various lengths (left) and coupling constants (right). The
BY equations describes the system more accurately at large and small
volumes and small coupling constants.}

\label{fig:bacheckevenlength}
\end{figure}

Let us examine now the validity of the Bethe--Yang equation for this
potential. Recall, that the scattering coefficient is given by eq.
(\ref{eq:bahyp}). However, now the Bethe--Yang equation does not
hold exactly neither in its symmetric nor its antisymmetric form.
To see this, we calculated the difference between the wavenumbers
computed numerically and the wavenumbers given with the help of Lamé
coefficients for different coupling constants and lengths. In figure
\ref{fig:bacheckevenlength} we can see, that the difference between
the BY wavenumbers and the exact ground state momentum decreases for
small and large lengths and small coupling constants. It would be
very interesting to describe this difference in terms of the scattering
phase.

\section{Ruijsenaars--Schneider models and the truncated Hilbert space method}

Having analysed the non-relativistic limits we turn back to the investigations
of the RS models. We focus on the two particle case in the center
of mass frame, where the Schrödinger equation takes the form 
\begin{equation}
H_{\mathrm{red}}\psi(q)=\left(f_{+}(q)T^{+}f_{+}(-q)+f_{+}(-q)T^{-}f_{+}(q)\right)\psi(q)=2\mathrm{cosh}(\theta)\psi(q)\,.\label{eq:RS2pt}
\end{equation}
where $T^{+}f(q)=f(q+i)$ and $T^{-}f(q)=f(q-i).$

We first present the infinite-volume RS models, which are the rational
and hyperbolic systems. As the potential vanishes at the infinities
we have free scattering solutions (\ref{eq:freeRS}): 
\begin{equation}
\psi_{q\to-\infty}(q)=e^{i\theta q}+S(\theta)e^{-i\theta q}\quad;\quad\psi_{q\to\infty}(q)=\pm(e^{-i\theta q}+S(\theta)e^{i\theta q})
\end{equation}
Note that here $q$ is not the coordinate, rather the conjugate variable
to the rapidity. We present first the phase for the hyperbolic model
and obtain the rational case as its limit.

We then turn to the periodic versions of these potentials. Recall
that here the periodicity is in $q$, not in the physical coordinate.
The real coordinate of the particle is $x=\frac{q}{\cosh\theta}$,
implying that the effective spatial periodicity depends on the particle\textquoteright s
momentum. Consequently, the notion of a periodic volume loses its
simple geometric meaning in $x$, and manifests itself only in $q$-space.
This feature is closely related to the relativistic invariance of
the model, which is preserved even though the system is defined in
a finite volume.

An approximate spectrum can be obtained by the Bethe-Yang equation
\begin{equation}
e^{i\theta L}S(\theta)=\pm1\label{eq:BAtheta}
\end{equation}
which quantises the rapidities and leads to energies via $E=2\cosh\theta$.

The exact finite volume solution is available for the trigonometric
case, which we present explicitly. We compare the result to the Bethe--Yang
equation and see that it is exact. As the relevant analytical solution
of the elliptic model for general coupling is not known we adapt the
truncated Hilbert space method to difference equations and benchmark
the result against the explicitly known integer $\lambda$ and trigonometric
cases. Finally, we use this method to obtain the spectrum of the generic
elliptic model.

In adapting the truncated Hilbert space method we split our Hamiltonian
into the already introduced free part and a perturbation 
\begin{equation}
H_{\mathrm{red}}=H_{\mathrm{free}}+H_{\mathrm{pert}}\,.
\end{equation}
where the perturbed Hamiltonian is given by
\begin{equation}
H_{\mathrm{pert}}=\left(f_{+}(q)f_{+}(-q-i)-1\right)T^{+}+\left(f_{+}(-q)f_{+}(q-i)-1\right)T^{-}\,.
\end{equation}
The basis introduced in the non-relativistic setting in the variable
$q$ is an eigenbasis of the free Hamiltonian 
\begin{equation}
H_{\mathrm{free}}\Phi_{n}^{(\pm)}=(T^{+}+T^{-})\Phi_{n}^{(\pm)}=2\cosh\left(\frac{2\pi}{L}n\right)\Phi_{n\,.}^{(\pm)}
\end{equation}
We then calculate the matrix elements of the perturbing Hamiltonian.
Since it is parity invariant we can work separately in the even and
odd subspaces: 
\begin{equation}
H_{\mathrm{pert},kl}^{(\pm)}=\langle\Phi_{k}^{(\pm)}|H_{\mathrm{pert}}|\Phi_{l}^{(\pm)}\rangle
\end{equation}
We now truncate our Hilbert space by keeping elements below the cutoff
$n\leq N$ and diagonalize the truncated reduced Hamiltonian $H_{\mathrm{red}}$.
By increasing $N$ we get a better approximation.

\subsection{The hyperbolic RS model}

The hyperbolic model was examined in details in \cite{ruijsenaars2001sine}.
For two particles in the center of mass frame the Schrödinger equation
takes the form (\ref{eq:RS2pt}) with the ``potential'' 
\begin{equation}
f_{+}^{\mathrm{hyp}}(q)=\sqrt{\frac{\sinh\left(d(q+i\lambda)\right)}{\sinh(dq)}}\,.
\end{equation}
As the potential vanishes at the infinities we should have free scattering
solutions (\ref{eq:freeRS}) and the aim is to determine the scattering
phase, which we denote by $S^{\mathrm{hyp}}(\theta)$. This job was
done by Ruijsenaars observing that the Schrödinger equation has also
a dual analogue \cite{ruijsenaars2001sine}, see also \cite{ruijsenaars1990finite,Ruijsenaars1999}.
This implied a difference equation directly for $S^{\mathrm{hyp}}(\theta)$
in $\theta$ \cite{Ruijsenaars:1997aqs}, which was analogous to the
unitarity and crossing requirement of the soliton-soliton scattering
phase and had the minimal solution
\begin{equation}
S^{\mathrm{hyp}}(\theta)=-\exp\left(2i\int_{0}^{\infty}\frac{dy}{y}\frac{\sinh(d(1-\lambda)y)\sinh(\pi y-d\lambda y)}{\sinh(dy)\sinh(\pi y)}\sin(y\theta)\right)\,.\label{eq:rshypscatter}
\end{equation}
This solution can be directly related to the sine-Gordon expression
by setting 
\begin{equation}
d=\frac{\pi p}{2}\quad;\quad\lambda=\frac{1}{p}
\end{equation}
In these calculations we set the speed of light to $c=1$ from the
beginning. If instead we restore its form and analyse its $c\to\infty$
limit we can recover the scattering phase (\ref{eq:bahyp}) of the
hyperbolic CMS model.

\subsection{The rational RS model}

The rational RS model is the $d\rightarrow0$ limit of the hyperbolic
model and is defined by the Schrödinger equation (\ref{eq:RS2pt}),
where the rational function $f_{+}^{\mathrm{rat}}$ is the double-degenerate
limit of the elliptic function:
\begin{equation}
f_{+}^{\mathrm{rat}}=\sqrt{\frac{q+i\lambda}{q}}\,.
\end{equation}
The scattering phase of the rational model can be obtained in the
limit $d\to0$ of the hyperbolic one (\ref{eq:rshypscatter}) $S^{\mathrm{rat}}(z)=\lim_{d\rightarrow0}S^{\mathrm{hyp}}(z)$,
which takes a particularly simple form:
\begin{equation}
\begin{gathered}S^{\mathrm{rat}}(\theta)=-\lim_{d\rightarrow0}\exp\left(2i\int_{0}^{\infty}\frac{dy}{y}\frac{\sinh\left((1-\lambda)y\right)\sinh(\pi y/d-\lambda y)}{\sinh y\sinh(\pi y/d)}\sin(y\theta/d)\right)=\\
=-\exp\left(2i(1-\lambda)\int_{0}^{\infty}\frac{dy}{y}\sin(y\theta)\right)=\exp\left(-i\pi\lambda\,\mathrm{sgn}(\theta)\right)\,,
\end{gathered}
\end{equation}
Clearly, the rational scattering coefficient does not depend on the
magnitude of the rapidity only on its sign. It is actually the same
as its non-relativistic limit, which is the rational CMS model.

\subsection{The trigonometric RS model}

The trigonometric RS model is the periodic version of the rational
model and is defined in the two-particle sector in the center of mass
frame by the Schrödinger equation(\ref{eq:RS2pt}) with the potential
function 
\begin{equation}
f_{+}^{\mathrm{trig}}(q)=\left(\frac{\sin\left(\pi(q+i\gamma)/L\right)}{\sin(\pi q/L)}\right)^{1/2}\,,\label{eq:trigrsfunc}
\end{equation}
This model yields back the rational RS model for $L\rightarrow\infty$.
This is the relativistic version of the trigonometric CMS model, and
can be solved analytically in a similar method. We are going to present
the analytical solutions first, and then we implement the truncated
Hilbert space method for difference equations and compare the results.

\subsubsection{Analytical solution of the trigonometric RS model}

Let us look for even solutions. First observe \cite{van1996diagonalization,arutyunov2019elements}
that the ground state function 
\begin{equation}
\psi_{0}(q)=\sqrt{\prod_{k=0}^{\infty}\frac{e^{i\frac{2\pi}{L}q}-e^{-\frac{2\pi}{L}k}}{e^{i\frac{2\pi}{L}q}-e^{-\frac{2\pi}{L}(k+\lambda)}}\frac{e^{-i\frac{2\pi}{L}q}-e^{-\frac{2\pi}{L}k}}{e^{-i\frac{2\pi}{L}q}-e^{-\frac{2\pi}{L}(k+\lambda)}}}\,
\end{equation}
solves our system and gives the ground state energy
\begin{equation}
E_{0}=2\cosh\left(\frac{\pi}{L}\lambda\right)\,.
\end{equation}
To get excited states we use a similar method to the trigonometric
Calogero-{}-Moser system: again, we look for solutions in the form
$\psi=\psi_{0}\tilde{\psi}$. After substituting and dividing by the
vacuum state we get the reduced equation
\begin{equation}
H_{\mathrm{red}}\tilde{\psi}(q)=\frac{\sin\left(\frac{\pi}{L}(q+i\lambda)\right)}{\sin\left(\frac{\pi}{L}q\right)}\tilde{\psi}(q+i)+\frac{\sin\left(\frac{\pi}{L}(q-i\lambda)\right)}{\sin\left(\frac{\pi}{L}q\right)}\tilde{\psi}(q-i)=E\tilde{\psi}(q)\,.
\end{equation}
The Hamiltonian $H_{\mathrm{red}}$ acts on the even basis spanned
by $\Phi_{n}^{(+)}$ as: 
\begin{equation}
\begin{gathered}H_{\mathrm{red}}\Phi_{n}^{(+)}=2\cosh\bigl(\frac{\pi}{L}(2n+\lambda)\bigr)(\Phi_{n}^{(+)}+\Phi_{n-1}^{(+)}+\dots+\frac{1}{2}\Phi_{0}^{(+)})-\\
-2\cosh\bigl(\frac{\pi}{L}(2n-\lambda)\bigr)(\Phi_{n-1}^{(+)}+\dots+\frac{1}{2}\Phi_{0}^{(+)})\,.
\end{gathered}
\end{equation}
implying that $\{\Phi_{0}^{(+)},\Phi_{1}^{(+)},\dots,\Phi_{n}^{(+)}\}$
is an invariant subspace and its elements form an upper triangular
matrix. The eigenvalues of the Hamiltonian are the diagonal elements
\begin{equation}
E_{n}=2\cosh\left(\theta_{n}\right)\,,\quad\theta_{n}=\frac{\pi}{L}\left(2n+\lambda\right)\,.\label{eq:evenrswavenumtrig}
\end{equation}
and the eigenvectors can be calculated iteratively.

The odd ground state wavefunction can be written in terms of the even
one as
\begin{equation}
\psi_{0}^{\mathrm{odd}}=\psi_{0}(q)\cos\left(\frac{\pi}{L}q\right)\mathrm{sgn}(q)\,
\end{equation}
where the continuity follows from $\psi_{0}(0)=0$. The corresponding
eigenvalue is 
\begin{equation}
E_{0}^{\mathrm{odd}}=2\cosh\left(\frac{\pi}{L}(1+\lambda)\right)\,.
\end{equation}
We can seek for solutions in the product form $\psi^{\mathrm{odd}}=\psi_{0}^{\mathrm{odd}}\tilde{\psi}_{\mathrm{odd}}$.
After dividing the Schrödinger equation by the odd vacuum state we
get
\begin{equation}
\begin{gathered}\frac{\sin\left(\frac{\pi}{L}(q+i\lambda)\right)}{\sin\left(\frac{\pi}{L}q\right)}\cos\left(\frac{\pi}{L}(q+i)\right)\tilde{\psi}_{\mathrm{odd}}(q+i)+\\
+\frac{\sin\left(\frac{\pi}{L}(q-i\lambda)\right)}{\sin\left(\frac{\pi}{L}q\right)}\cos\left(\frac{\pi}{L}(q-i)\right)\tilde{\psi}_{\mathrm{odd}}(q-i)=E^{\mathrm{odd}}\tilde{\psi}_{\mathrm{odd}}(q)\cos\left(\frac{\pi}{L}q\right)\,.
\end{gathered}
\end{equation}
Using the even basis (\ref{eq:simplebasis}) we obtain 
\begin{equation}
\begin{gathered}H_{\mathrm{red}}\Phi_{n}^{(+)}=2\cosh\left(\frac{\pi}{L}\left((2n+1)+\lambda\right)\right)\Phi_{n}^{(+)}+\\
+\left(2\cosh\left(\frac{\pi}{L}\left((2n+1)+\lambda\right)\right)-2\cosh\left(\frac{\pi}{L}\left((2n-1)-\lambda\right)\right)\right)(\Phi_{n-2}^{(+)}+\Phi_{n-4}^{(+)}+\dots)+\\
+\left(2\cosh\left(\frac{\pi}{L}\left((2n-1)+\lambda\right)\right)-2\cosh\left(\frac{\pi}{L}\left((2n+1)-\lambda\right)\right)\right)(\Phi_{n-1}^{(+)}+\Phi_{n-3}^{(+)}+\dots)\,,
\end{gathered}
\end{equation}
so the action of $H_{\mathrm{red}}$ on $\{\Phi_{0}^{(+)},\dots,\Phi_{n}^{(+)}\}$
yields an upper triangular matrix with the following eigenvalues:
\begin{equation}
E_{n}^{\mathrm{odd}}=2\cosh\left(\theta_{n}\right)\,,\quad\theta_{n}=\frac{\pi}{L}\left(2n+1+\lambda\right)\,,\label{eq:oddrswavenumtrig}
\end{equation}
which complete our even-state rapidities. Clearly we get back both
the odd and even momenta of the trigonometric CM model. The difference,
however is, that in the RS case the rapidities agree with wavenumbers
and not the momenta. This is related to the fact that the wave function
has the form $e^{i\theta q}$ and not $e^{ipx}$.

\subsubsection{Comparison to BY equations}

Similarly to the trigonometric CM model, substituting the quantised
rapidities (\ref{eq:evenrswavenumtrig}) ( for even states) and (\ref{eq:oddrswavenumtrig})
(for odd states) into eq. (\ref{eq:BAtheta}) we can see, that the
BY equations hold for both the even and odd cases, the latter giving
the antisymmetric BY equations. Again, as in case of the trigonometric
Calogero--Moser system, we can conclude that the Bethe--Yang equations
hold for any length $L$.

\subsection{Numerical calculations for the trigonometric RS model}

We proceed to solve the trigonometric RS model numerically. We follow
the generic procedure of splitting the Hamiltonian into a free and
interacting part and calculate the matrix elements of the perturbation
directly. We start with the odd matrix elements 

\begin{equation}
=\begin{gathered}H_{\mathrm{pert},kl}^{(-)}=\langle\Phi_{k}^{(-)}|H_{\mathrm{pert}}|\Phi_{l}^{(-)}\rangle=\\
\frac{2}{L}\int_{-L/2}^{L/2}dq\sin\left(\frac{2\pi kq}{L}\right)\left(f_{+}^{\mathrm{trig}}(q)f_{+}^{\mathrm{trig}}(-q-i)-1\right)\sin\left(\frac{2\pi l(q+i)}{L}\right)+\\
+\frac{2}{L}\int_{-L/2}^{L/2}dq\sin\left(\frac{2\pi kq}{L}\right)\left(f_{+}^{\mathrm{trig}}(-q)f_{+}^{\mathrm{trig}}(q-i)-1\right)\sin\left(\frac{2\pi l(q-i)}{L}\right)\,.
\end{gathered}
\label{eq:pertoddham}
\end{equation}
The matrix elements of the full Hamiltonian can be calculated as
\begin{equation}
H_{\mathrm{red},kl}^{(-)}=H_{\mathrm{free},kl}^{(-)}+H_{\mathrm{pert},kl}^{(-)}\,,\label{eq:rsfullmatrixodd}
\end{equation}
which, after truncation, can be diagonalized numerically. This provides
approximate eigenvalues and eigenvectors. By increasing the truncation
levels the method nicely converges.

Although the similar procedure works for even states the singular
behaviour of the potential at the origin $\vert q\vert^{-\frac{1}{2}}$
prevents the method from converging well. So instead we introduced
a new basis:
\begin{equation}
\Phi_{k}=\sqrt{\frac{2}{L}}\sqrt{\sin^{2}(\pi q/L)}\cos\left(\frac{2\pi}{L}kq\right)\,.\label{eq:evenbasisrs}
\end{equation}
As this basis is not orthonormal we have to proceed as we did in the
case of the CMS models. First, we calculate the matrix elements of
the free Hamiltonian in this basis:
\begin{equation}
\begin{gathered}H_{\mathrm{free},kl}^{\mathrm{even}}=\langle\Phi_{k}|H_{\mathrm{free}}|\Phi_{l}\rangle=\cosh\left(\frac{\pi}{L}\right)\cosh\left(\frac{2\pi}{L}k\right)\delta_{kl}-\frac{1}{2}\cosh\left(\frac{\pi}{L}(2k-1)\right)\delta_{k,l+1}-\\
-\frac{1}{2}\cosh\left(\frac{\pi}{L}(2k+1)\right)\delta_{k,l-1}+\cosh\left(\frac{\pi}{L}\right)\delta_{k0}\delta_{l0}-\frac{1}{2}\cosh\left(\frac{\pi}{L}\right)(\delta_{k0}\delta_{l1}+\delta_{l0}\delta_{k1})\,.
\end{gathered}
\label{eq:rsevenfreematrixelements}
\end{equation}
The matrix elements of the perturbed Hamiltonian must be calculated
numerically. It has the following matrix elements:
\begin{equation}
\begin{gathered}H_{\mathrm{pert},kl}^{\mathrm{even}}=\langle\Phi_{k}|H_{\mathrm{pert}}|\Phi_{l}\rangle=\\
=\int_{-L/2}^{L/2}dq\ \Phi_{k}(q)\left(f_{+}^{\mathrm{trig}}(q)f_{+}^{\mathrm{trig}}(-q-i)-1\right)\Phi_{k}(q+i)+\\
\int_{-L/2}^{L/2}dq\ \Phi_{k}(q)\left(f_{+}^{\mathrm{trig}}(-q)f_{+}^{\mathrm{trig}}(q-i)-1\right)\Phi_{k}(q-i)\,.
\end{gathered}
\label{eq:pertevenham}
\end{equation}
Again, we use for $f_{+}$ the function given by eq. (\ref{eq:trigrsfunc}).
In this case, the basis elements are not orthonormal, so we need to
calculate the inverse of the matrix obtained from the inner products.
The inner product in this basis can be given as
\begin{equation}
\langle\Phi_{k}|\Phi_{l}\rangle=\frac{1}{2}\delta_{kl}-\frac{1}{4}(\delta_{k,l-1}+\delta_{k,l+1})+\frac{1}{2}\delta_{k0}\delta_{l0}-\frac{1}{4}(\delta_{k0}\delta_{l1}+\delta_{l0}\delta_{k1})\,.
\end{equation}
To determine the eigenvalues and eigenvectors, we need to calculate
the inverse of the inner product. If we act on the truncated basis
$\{\Phi_{0},\Phi_{1},\dots,\Phi_{N}\}$ we get the following inverse
inner product:
\begin{equation}
\left(\Phi_{kl}\right)^{-1}=\begin{pmatrix}N+1 & 2N & 2N-2 & 2N-4 & \dots & 2\\
2N & 4N & 4N-4 & 4N-8 & \dots & 4\\
2N-2 & 4N-4 & 4N-4 & 4N-8 & \dots & 4\\
2N-4 & 4N-8 & 4N-8 & 4N-8 & \dots & 4\\
\vdots & \vdots & \vdots & \vdots & \ddots & \vdots\\
2 & 4 & 4 & 4 & \dots & 4
\end{pmatrix}\,.\label{eq:iiprs}
\end{equation}
The matrix elements of the full Hamiltonian can be calculated as
\begin{equation}
H_{\mathrm{red},kl}^{\mathrm{even}}=\left(\Phi_{kj}\right)^{-1}\left(H_{\mathrm{free},jl}^{\mathrm{even}}+H_{\mathrm{pert},jl}^{\mathrm{even}}\right)\,.\label{eq:rsfullmatrixeven}
\end{equation}
We have to calculate the eigenvalues and eigenvectors of $(H_{\mathrm{red}}^{\mathrm{even}})_{kl}$
to get the even eigenstates and their energies and wavenumbers numerically.
After getting the odd and even eigenvalues $E_{k}$ we can calculate
the corresponding quantised rapidities as
\begin{equation}
\theta_{n}=\arccos(E_{k}/2)\,.
\end{equation}

\begin{figure}[H]
\begin{centering}
\includegraphics[scale=0.75]{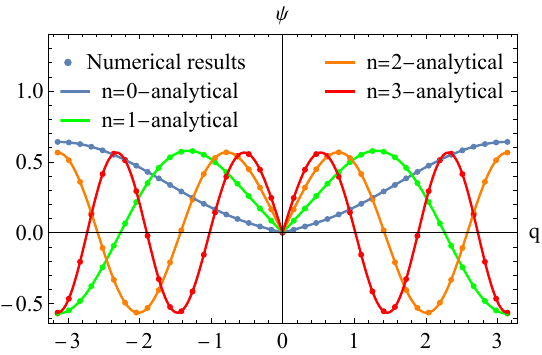}\includegraphics[scale=0.75]{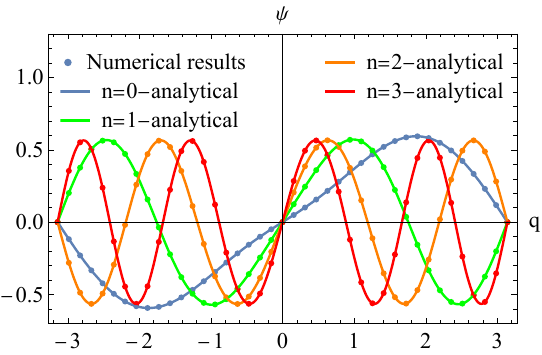}
\par\end{centering}
\caption{The first four even (left) and odd (right) eigenstates with their
eigenvalues calculated analytically and numerically and normalised
for $L=2\pi$, $\lambda=2.3$, $N=10$.}
\label{fig:rstrigodd}
\end{figure}

We plotted the first four odd and even eigenfunctions for $L=2\pi$
and $N=10$ in figure \ref{fig:rstrigodd}. The numerical states and
eigenvalues correspond nicely to the analytical states and their eigenvalues,
which satisfy the Bethe--Yang equations.

\subsection{The elliptic RS model}

Finally, let us turn our attention to the elliptic RS model, which
is the finite-volume version of the hyperbolic RS model. We want to
solve the Schrödinger equation (\ref{eq:RS2pt}) with the elliptic
potential $f_{+}^{\mathrm{ell}}$ written in terms of the Weierstrass
sigma functions \cite{Ruijsenaars:1986pp}:
\begin{equation}
f_{+}^{\mathrm{ell}}(q)=\sqrt{\frac{\sigma(q+i\lambda,L,i\pi/d)}{\sigma(q,L,i\pi/d)}}\,.
\end{equation}
Although the sigma functions are quasi-periodic with the relation
\begin{equation}
\sigma(q+L)=-e^{2\eta_{1}(q+L/2)}\sigma(q)\,,\label{eq:sigmaquasiperiod}
\end{equation}
where $\eta_{1}$ is the Weierstrass zeta-function at $L/2$, the
Hamiltonian $H_{\mathrm{red}}^{\mathrm{ell}}$ itself is periodic.
To see this, we can write it in the explicit form:
\begin{equation}
H_{\mathrm{red}}^{\mathrm{ell}}=g_{+}(q)T^{+}+g_{-}(q)T^{-}\,,
\end{equation}
where
\begin{equation}
g_{\pm}(q)=\sqrt{\frac{\sigma(q\pm i\lambda,L,i\pi/d)}{\sigma(q,L,i\pi/d)}}\sqrt{\frac{\sigma(q\mp i\lambda\pm i,L,i\pi/d)}{\sigma(q\pm i,L,i\pi/d)}}
\end{equation}
and observe the periodicity $g_{\pm}(q+L)=g_{\pm}(q)$.

It is useful to see how the hyperbolic theory is recoverd in the limit
$L\rightarrow\infty$. Since the sigma function behaves as \cite{Ruijsenaars:1986pp}:
\begin{equation}
\sigma(q,\infty,id)=\frac{\sinh dq}{d}\exp\left(-\frac{d^{2}}{6}q^{2}\right)\,.
\end{equation}
we can obtain the limiting cases
\begin{equation}
g_{\pm}(q)\xrightarrow{L\rightarrow\infty}\exp\left(\frac{d^{2}}{6}\lambda(\lambda-1)\right)\sqrt{\frac{\sinh(d(q\pm i\lambda))}{\sinh(dq)}}\sqrt{\frac{\sinh(d(q\mp i\lambda\pm i))}{\sinh(d(q+i))}}
\end{equation}
which leads to the following relations of the Hamiltonians 
\begin{equation}
H_{\mathrm{red}}^{\mathrm{ell}}\xrightarrow{L\rightarrow\infty}\exp\left(\frac{d^{2}}{6}\lambda(\lambda-1)\right)\left(f_{+}^{\mathrm{hyp}}(q)T^{+}f_{+}^{\mathrm{hyp}}(-q)+f_{+}^{\mathrm{hyp}}(-q)T^{-}f_{+}^{\mathrm{hyp}}(q)\right)\,,
\end{equation}
They thus agree up to a multiplicative factor 
\begin{equation}
H_{\mathrm{red}}^{\mathrm{ell}}\xrightarrow{L\rightarrow\infty}CH_{\mathrm{red}}^{\mathrm{hyp}}\,,\quad C=\exp\left(\frac{d^{2}}{6}\lambda(\lambda-1)\right)\label{eq:multfact}
\end{equation}
This means that the finite volume version of the hyperbolic model
is actually $C^{-1}H_{\mathrm{red}}^{\mathrm{ell}}$. In order to
take care of this difference and support the correct comparison with
the Bethe--Yang equations we parametrize the energy eigenvalues in
the elliptic Schrödinger equation in the following form:
\begin{equation}
\left(f_{+}^{\mathrm{ell}}(q)T^{+}f_{+}^{\mathrm{ell}}(-q)+f_{+}^{\mathrm{ell}}(-q)T^{-}f_{+}^{\mathrm{ell}}(q)\right)\psi(q)=2C\cosh(\theta)\psi(q)\,.
\end{equation}
The perturbative solution was constructed in \cite{langmann2022construction}
at generic coupling. Here we determine the eigenvalues and eigenvectors
numerically using the non-perturbative truncated Hilbert space method.
We benchmark the solution against the explicitly known solution for
$\lambda=2$, which we recall now from \cite{Ruijsenaars2000}.

\subsubsection{Analytical solution for $\lambda=2$}

\begin{figure}
\begin{centering}
\includegraphics[scale=0.75]{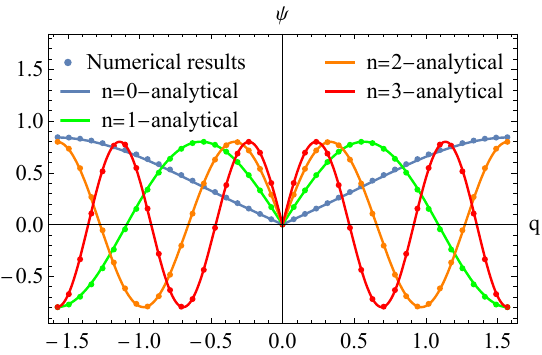}\hspace{1cm}\includegraphics[scale=0.75]{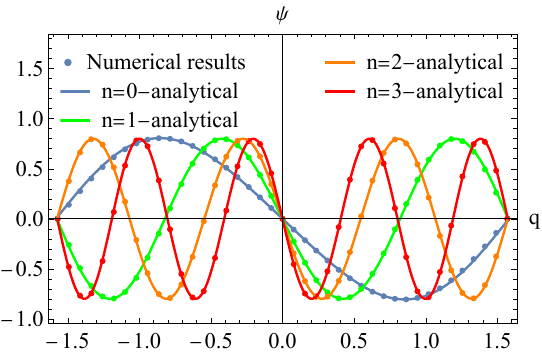}
\par\end{centering}
\caption{The first four even (left) and odd (right) eigenstates of the elliptic
RS model with their eigenvalues calculated analytically and numerically
and normalised for $L=\pi$, $\lambda=2$, $N=10$.}
\label{fig:lambda2rsell}
\end{figure}

Similarly to the elliptic Calogero-Moser system the elliptic RS model
can be solved using transcendental functions \cite{Ruijsenaars2000}.
We present, again the simplest solution for the elliptic RS model
for $\lambda=2$. First, we consider the following functions which
solve our elliptic difference equation:

\begin{equation}
\mathcal{F}(q,y)=\frac{\sigma(q+z)}{\sigma(q+i)\sigma(q-i)}\exp(2\text{\ensuremath{\pi}}iq/L+iqy)
\end{equation}
with the following energy eigenvalue:
\begin{equation}
E=\frac{1}{\sigma(q)}\bigl(\sigma(q+i)\exp(2\pi/L+y)\frac{\sigma(q-i+z)}{\sigma(q+z)}+\sigma(q-i)\exp(-2\pi/L-y)\frac{\sigma(q+i+z)}{\sigma(q+z)}\bigr)\,.\label{eq:ellipticenergy}
\end{equation}
This energy depends on $q$ for generic $y$ and $z$ values. However,
it is an elliptic function with the periods $L$ and $\pi/d$, so
it is constant if the residue of one of its two simple poles vanish
\cite{Ruijsenaars2000,whittaker2020course}. The pole at $q=0$ vanishes
if
\begin{equation}
y=-\frac{2\pi}{L}-\frac{1}{2}\ln\frac{\sigma(z-i)}{\sigma(z+i)}\,.
\end{equation}
In addition of the condition for the constant energy, we need $\mathcal{F}$
to be periodic in $L$. If we seek for the coefficient $z$ in the
form of $z=i\alpha_{n}$ we get the following condition:
\begin{equation}
4\eta_{1}\alpha_{n}-L\ln\frac{\sigma(i\alpha_{n}-i)}{\sigma(i\alpha_{n}+i)}=2(n+2)\pi\,.
\end{equation}
Substituting back the values of $z=i\alpha_{n}$ into (\ref{eq:ellipticenergy})
we get back the energies calculated for even states for even $n$
values starting with $n=0$ and the energies of odd states for odd
values starting with $n=1$. The following eigenfunctions coincide
with the numerical odd and even eigenfunctions, respectively \cite{Ruijsenaars2000}:
\begin{equation}
\begin{cases}
\psi_{n}=\mathcal{F}(q,\alpha_{n})-\mathcal{F}(-q,\alpha_{n}) & n\,\mathrm{odd}\\
\psi_{n}=\mathrm{sgn}(q)(\mathcal{F}(-q,\alpha_{n})-\mathcal{F}(q,\alpha_{n})) & n\,\mathrm{even}
\end{cases}
\end{equation}
The analytical and numerical wavefunctions are shown in figure \ref{fig:lambda2rsell}
for the first four odd and even states for $\lambda=2$ and $L=\pi$.

\subsubsection{Numerical solution of the elliptic RS model}

We proceed with the numerical solution for generic coupling as in
the case of the trigonometric RS model. We start with the odd subspace
spanned by $\Phi_{n}^{(-)}$ and calculate the matrix elements of
the perturbing Hamiltonian numerically
\begin{equation}
\begin{gathered}H_{\mathrm{pert},kl}^{(-)}=\langle\Phi_{k}^{(-)}|H_{\mathrm{pert}}|\Phi_{l}^{(-)}\rangle=\int_{-L/2}^{L/2}dq\,\Phi_{k}^{(-)}(q)\left(f_{+}^{\mathrm{ell}}(q)f_{+}^{\mathrm{ell}}(-q-i)-1\right)\Phi_{l}^{(-)}(q+i)+\\
+\int_{-L/2}^{L/2}dx\,\Phi_{k}^{(-)}(q)\left(f_{+}^{\mathrm{ell}}(-q)f_{+}^{\mathrm{ell}}(q-i)-1\right)\Phi_{k}^{(-)}(q-i)
\end{gathered}
\end{equation}
In the even case we could use the even basis $\Phi_{n}^{(+)},$ however,
the convergence problem due to the $\vert q\vert^{-1/2}$ behaviour
persists, so we switched to the basis $\Phi_{n}$. After calculating
the perturbed matrices, we calculate again the eigenvalues and eigenvectors
of the truncated Hamiltonians (\ref{eq:rsfullmatrixodd}) and (\ref{eq:rsfullmatrixeven}).
From any eigenvalue we can calculate the corresponding rapidity $\theta_{n}$.

\begin{figure}[H]
\begin{centering}
\includegraphics[scale=0.75]{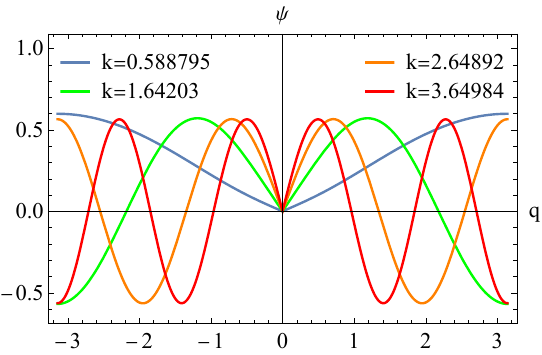}\includegraphics[scale=0.75]{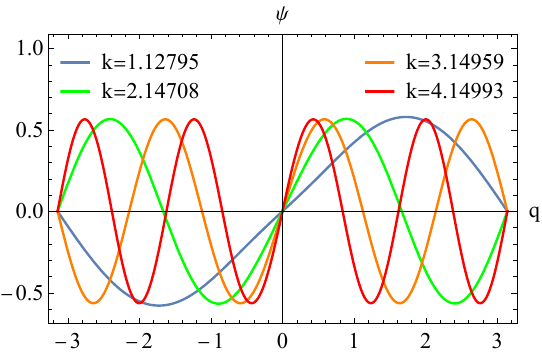}
\par\end{centering}
\caption{Numerically calculated even (left) and odd (right) states for the
elliptic RS model. We used the parameters $L=2\pi$, $d=5\pi/16$,
$\lambda=8/5$ and $N=10$.}
\label{fig:rsellodd}
\end{figure}

We plotted the first four even and odd eigenstates for $L=2\pi$ and
$d=5\pi/16$ in figure \ref{fig:rsellodd}. These parameters correspond
to $p=5/8$ in the sine-Gordon model.

\subsubsection{Comparison to Bethe--Yang equations}

\begin{figure}[H]
\begin{centering}
\includegraphics[width=7cm]{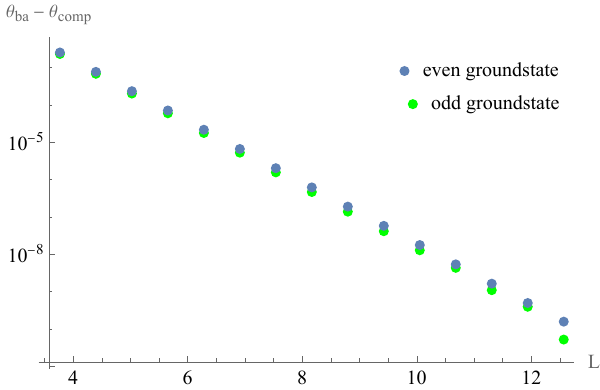}\includegraphics[width=7cm]{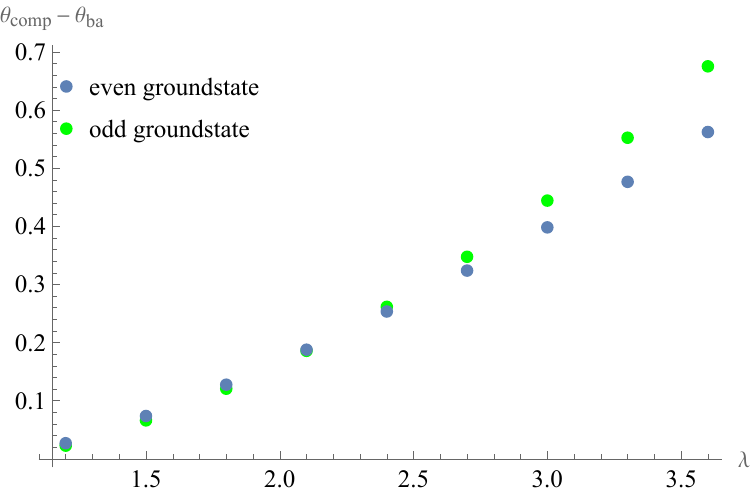}
\par\end{centering}
\caption{Validity of the BY equations for the even and odd ground states at
various lengths (left) and coupling constants (right). The Bethe--Yang
equation describes the system more accurately at large volumes and
small coupling constants.}

\label{fig:rsbaoddlength}
\end{figure}

As in the case of the elliptic CMS model, the BY equations does not
hold exactly for the calculated rapidities. To see this, we calculated
the difference between the numerically computed quantised rapidities
and their BY analogues for the even and odd ground states for different
lengths and coupling constants $\lambda$. We can see in figure \ref{fig:rsbaoddlength},
that the difference between the rapidities is smaller for larger lengths
and smaller couplings, which is also true for the elliptic CMS models.

\section{Comparing the quantum elliptic RS model to the finite volume quantum
sG theory}

Let us point out an important difference between quantum field theories
and quantum many body systems. The vacuum in a quantum field theory
is a complicated state, which is populated by particle--anti-particle
pairs, while in quantum mechanics it is empty. When a QFT is obtained
as a thermodynamic limit of a many body system, the vacuum is typically
a strongly interacting state. The finite volume sine-Gordon theory,
for instance, can be obtained as the continuum scaling limit of the
staggered XXZ spin chain \cite{Destri:1994bv}. The finite volume
sine-Gordon vacuum corresponds to the anti-ferromagnetic XXZ vacuum,
with infinitely many interacting real Bethe roots. As a consequence,
the ground state energy has a non-trivial volume dependence. In the
following we compare the energy difference between a two soliton state
and the vacuum to the spectrum of the elliptic RS model. We normalise
the elliptic spectrum, such that at large volumes it reproduces the
hyperbolic model (\ref{eq:multfact}).

\subsection{Free case}

First, let us examine the free theories. This corresponds to $p=1$
in the sine-Gordon theory, when the scattering phase vanishes. The
quantization condition for two moving free fermions is 
\begin{equation}
ML\sinh\theta=2\pi n+\pi
\end{equation}
where $n$ is an integer. Taking $M=1$, the corresponding energy
difference to the vacuum is 
\begin{equation}
E=2\cosh\theta\,.
\end{equation}
The free RS model is given by $d=\frac{\pi}{2}$ and $\lambda=1$.
The Bethe ansatz is exact and quantises the rapidities as 
\begin{equation}
\theta L=2\pi n+\pi
\end{equation}
while the energy is given by
\begin{equation}
E_{\mathrm{RS}}=2\cosh\theta\,.
\end{equation}
This means, that the spectrum of the two free theories coincide only
at very large lengths, where we can take the approximation $\sinh\theta\approx\theta$.
This is related to the fact that the wave function in the sine-Gordon
case is $e^{ipx}$, while in the RS case it is $e^{i\theta q}$. This
former expression is required by the correct dispersion relation.

\subsection{Interacting case}

Next, we compare the spectrum of the elliptic RS model and the sG
model. We do the comparison only numerically. We should keep in mind,
that after calculating the eigenvalues of the RS model with the truncated
Hilbert space method, we should multiply them with the inverse of
the multiplicative factor given in (\ref{eq:multfact})
\begin{equation}
C^{-1}=\exp\left(-\frac{d^{2}}{6}\lambda(\lambda-1)\right)=\exp\left(-\frac{\pi^{2}}{24}(1-p)\right)
\end{equation}
 to get the correct spectrum.

At larger volumes we expect that both the sine-Gordon theory and the
RS model approximate the Bethe--Yang quantisation lines. However,
the two BY equations are different, which has its roots in the different
coordinate representations, similarly to the free case. Indeed, the
BY equations in the sine-Gordon theory are (\ref{eq:sGBY})
\[
e^{iL\sinh\theta}S(\theta)=1
\]
 In contrast, in the RS model they read as 
\[
e^{iL\theta}S(\theta)=1
\]
while the corresponding energies have the same form $E=2\cosh\theta$.
Thus already at the BY level we recognize differences, which is due
to the different forms of the asymptotic wave functions. This difference
could have dissappeared in the exact spectrum, but we can do this
comparision only numerically.

\begin{figure}
\begin{centering}
\includegraphics[width=7cm]{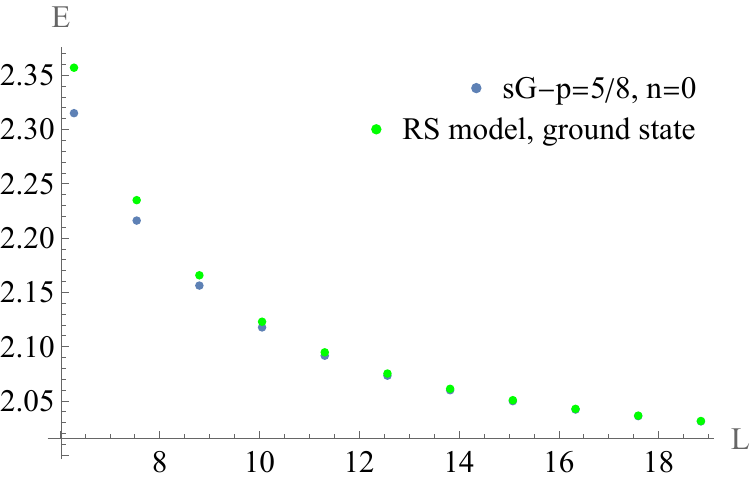}\hspace{1cm}\includegraphics[width=7cm]{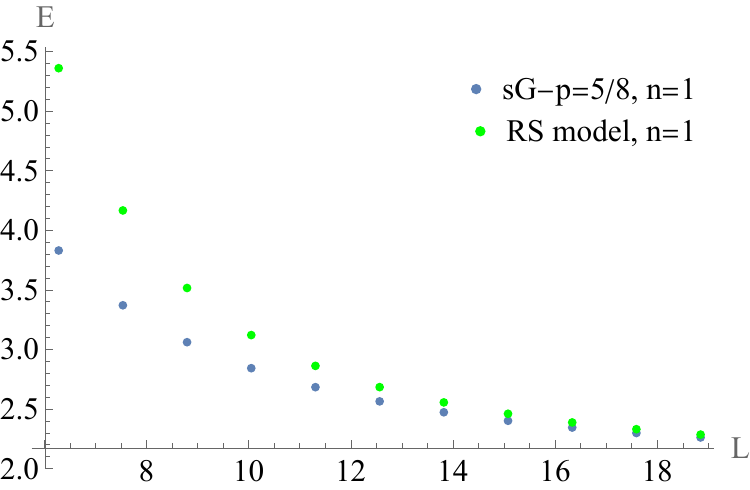}
\par\end{centering}
\caption{Ground state and first even excited state energies for the RS and
first and second corrected two-soliton energies for the sG models
for different system volumes. We used $p=5/8$.}

\label{fig:comp581}
\end{figure}

\begin{figure}
\begin{centering}
\includegraphics[width=7cm]{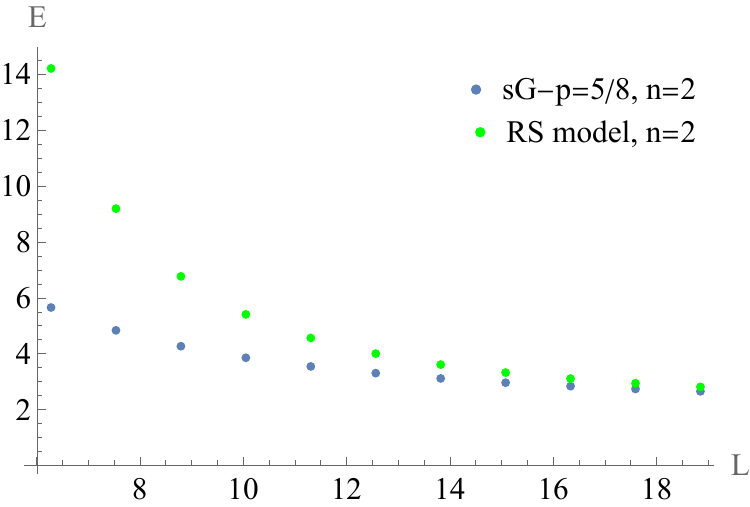}\hspace{1cm}\includegraphics[width=7cm]{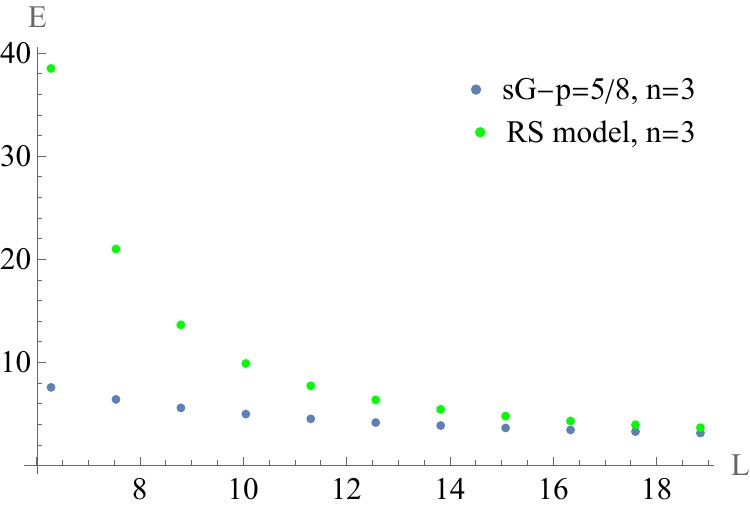}
\par\end{centering}
\caption{Second and third even excited state energies for the RS and third
and fourth corrected two-soliton energies for the sG models for different
system volumes. We used $p=5/8$.}

\label{fig:comp583}
\end{figure}

In figures \ref{fig:comp581} and \ref{fig:comp583} we compare the
spectra of the elliptic RS and the finite-volume sG model at $p=5/8$
for different lengths. In case of the sine-Gordon model, we subtracted
the vacuum energy $E_{0}$ from the soliton energies. As expected,
the two spectra approach each other for large lengths, where they
approximate the Bethe ansatz equations, and the difference between
the energies is smaller for smaller values of $n$.

We can clearly see on the figures that the two models' energies differ.
So the difference appears not only because they approach different
BY energies at large energies, but they are really different at finite
sizes. We investigated the cutoff $(N)$ depence of these results
and observed that errors were smaller than the dots on the figures
and decreased with increasing the cutoff. 

\section*{Conclusions}

In this paper we compared the finite-size spectrum of the sine-Gordon
quantum field theory with the spectrum of the elliptic Ruijsenaars--Schneider
model. The original motivation was to test whether the well-established
equivalence between the infinite-volume sine-Gordon theory and the
hyperbolic RS model extends to finite volume. Although the two models
share the same scattering matrix in infinite volume, finite-size effects
in relativistic QFTs are governed by the scattering phases not only
via the Bethe--Yang equations but also by vacuum polarization (Lüscher)
corrections. No analogous, generally accepted finite-volume framework
exists for quantum many-body systems, so we pursued a predominantly
numerical comparison.

Since there are no non-perturbative exact results for the elliptic
RS model, we adapted the non-perturbative truncated Hilbert-space
method to difference operators and benchmarked it in several cases
(integer coupling, the trigonometric limit), where analytical results
are available. Our numerical study---restricted to the two-particle
sector in the center-of-mass frame---shows that the two spectra differ
at finite volume. The differences can be traced to (i) the nontrivial
vacuum structure of the field theory (vacuum polarisation by particle--antiparticle
pairs) versus the trivial vacuum of the quantum many-body system,
and (ii) the distinct form of the free wavefunctions: in the QFT they
are plane waves in coordinate space, whereas in the RS models plane
waves live in the coordinate dual to rapidity.

In the large-volume limit both theories are governed at leading order
by the Bethe--Yang equations, but we find that subleading (finite-size)
corrections behave differently: the Bethe--Yang quantization is exact
in the rational/trigonometric correspondences, while in the hyperbolic/elliptic
case it is only asymptotic. This suggests that finite-size corrections
for many-body systems may nevertheless be expressible in terms of
scattering data (for example, via derivatives of the S-matrix), but
that the precise mechanism differs from the Lüscher/TBA machinery
of relativistic QFTs.

A natural next step is to develop an analytic description of these
finite-size corrections for quantum many-body systems (elliptic RS
and their non-relativistic limits). Clarifying whether and how such
corrections can be written solely in terms of the S-matrix would both
deepen the RS--sine-Gordon correspondence and broaden our understanding
of finite-volume effects in integrable quantum mechanics.

Promising next steps include deriving Lüscher-like formulae for many-body
systems and searching for a TBA-style formulation that captures the
elliptic finite-volume corrections.

\section*{Acknowledgements}

We are indebted to Rob Klabbers for the many useful discussions and
comments on the manuscript. We also thank János Balog and László Fehér
for reading the manuscript and the useful suggestions. ZB thanks the
Marc Kac center in Kraków for hospitality during the final stages
of this work. ZB was supported in part by a Priority Research Area
DigiWorld grant under the Strategic Programme Excellence Initiative
at the Jagiellonian University (Kraków, Poland). The research was
also supported by the grant K134946 of NKFIH.

\appendix

\section{Details of the elliptic CMS model}

In this Appendix we detail the calculations of the elliptic CMS model.
We map the differential equation to the Heun equation and investigate
its regularity properties. We then compare for integer $\lambda$s
the results to the Lame solution.

\subsection{Analytical solution}

We need to solve the following equation:
\begin{equation}
-\frac{\partial^{2}\psi(x)}{\partial x^{2}}+\left(\wp(x,L,\frac{i\pi}{d})+\frac{2d\eta_{3}}{i\pi}\right)\lambda(\lambda-1)\psi(x)=E\psi(x)\,,
\end{equation}
As we already shown, using functional identities with Jacobi elliptic
functions and changing the variable to $z=\sqrt{e_{1}-e_{3}}x$ we
obtain the differential equation
\begin{equation}
-\frac{\partial^{2}\psi(z)}{\partial z^{2}}+\frac{\lambda(\lambda-1)}{\text{sn}^{2}(z,m)}\psi(z)=\left(\frac{E}{e_{1}-e_{3}}-\frac{\lambda(\lambda-1)e_{3}}{e_{1}-e_{3}}-\frac{\lambda(\lambda-1)2d\eta_{3}}{i\pi(e_{1}-e_{3})}\right)\psi(z)\equiv h\psi(z)\,.\label{eq:sninverse-1}
\end{equation}
where the periodicity requirement is $\psi(z+\sqrt{e_{1}-e_{3}}L)=\psi(z)$.
This is not of the standard Lame form, but using the identity ${\rm sn}(z,m)^{-2}=m\,{\rm sn}(z+i\frac{\sqrt{e_{1}-e_{3}}\pi}{d},m)^{2}$
we can write 
\begin{equation}
-\frac{\partial^{2}\psi(z)}{\partial z^{2}}+m\lambda(\lambda-1)\text{sn}^{2}\biggl(z+i\frac{\sqrt{e_{1}-e_{3}}\pi}{d},m\biggr)\psi(z)=h\psi(z)
\end{equation}
We can introduce the shifted variable $\tilde{z}=z+i\frac{\sqrt{e_{1}-e_{3}}\pi}{d}$
but the periodicity requirement needs to be imposed at an imaginary
shifted postion $\tilde{z}-i\frac{\sqrt{e_{1}-e_{3}}\pi}{d}$. This
is very non-standard so we transform instead the equation (\ref{eq:sninverse-1})
into a Heun equation.

Let us start with the even solutions and write the wavefunction as
\begin{equation}
\psi(z)=|\mathrm{sn}(z)|^{\lambda}\tilde{\psi}(z)\,.
\end{equation}
and substitute $u=\mathrm{sn}^{2}(z)$ to get the form the Heun equation
\cite{ronveaux1995heun}:
\begin{equation}
\frac{\partial^{2}\tilde{\psi}(u)}{\partial u^{2}}+\left(\frac{\gamma}{u}+\frac{\delta}{u-1}+\frac{\epsilon}{u-a}\right)\frac{\partial\tilde{\psi}(u)}{\partial u}+\frac{\alpha\beta u-q}{u(u-1)(u-a)}\tilde{\psi}(u)=0\,
\end{equation}
with the parameters $a=\frac{1}{m}$, $q=\frac{\lambda^{2}(1+\frac{1}{m})-\frac{h}{m}}{4}$,
$\alpha=\frac{\lambda}{2}$, $\beta=\frac{\lambda+1}{2}$, $\gamma=\frac{1+2\lambda}{2}$,
$\delta=\frac{1}{2}$ and $\epsilon=\frac{1}{2}$. The solution of
the spectral problem then have the form
\begin{equation}
\psi(z)=|\mathrm{sn}(z)|^{\lambda}Hl\left(\frac{1}{m},\frac{\lambda^{2}(1+\frac{1}{m})-\frac{h}{m}}{4},\frac{\lambda}{2},\frac{\lambda+1}{2},\frac{1+2\lambda}{2},\frac{1}{2},\text{sn}^{2}(z)\right)\,.\label{eq:heunevensmall-1}
\end{equation}

We now investigate the regularity of the solution. Recall that the
general Heun equation has four singular points at $0$, $1$, $a$
and $\infty$. The general Heun function $Hl(a,q,\alpha,\beta,\gamma,\delta,u)$
is constructed to be regular at $u=0$, however, the singularities
at $1$, $a$ and $\infty$ persist in these solutions \cite{ronveaux1995heun}.
As in our case $a=\frac{1}{m}\geq1$ and $-1\leq\mathrm{sn}^{2}(z)\leq1$
the singularities at $a$ and $\infty$ should cause no concerns,
however, the solution remains singular around 1, which corresponds
to $\mathrm{sn}^{2}(\frac{\sqrt{e_{1}-e_{3}}L}{2})$, the point in
the middle of the periodicity strip. To understand to regularity requirement
we investigate the convergence properties of the power series expansion
around $u=0$. It is known \cite{ronveaux1995heun}, that the radius
of convergence of these solutions is usually $1$, and the solutions
are singular at 1, however the radius of convergence can be extended
to $\frac{1}{m}$ for certain values of $h$. Let us see, how the
requirement comes out.

First, we construct the generic solutions as a powever series \cite{ronveaux1995heun}:
\begin{equation}
Hl(z)=\sum_{r=0}^{\infty}c_{r}z^{r}\,.\label{eq:powseries-1}
\end{equation}
 Plugging black into the differential we obtain the following recursion
\cite{ronveaux1995heun}:
\begin{equation}
\begin{cases}
P_{r}c_{r-1}+S_{r}c_{r}+R_{r}c_{r+1}=0 & r\geq1\\
-qc_{0}+a\gamma c_{1}=0\,,
\end{cases}\label{eq:genrecursion-1}
\end{equation}
with
\begin{align}
P_{r} & =(r-1+\alpha)(r-1+\beta)\nonumber \\
S_{r} & =-q-r\left((r-1+\gamma)(1+a)+a\delta+\epsilon\right)\nonumber \\
R_{r} & =(r+1)(r+\gamma)a
\end{align}
where in our case
\begin{equation}
\begin{cases}
\begin{gathered}m(\lambda+2r-2)(\lambda+2r-1)c_{r-1}+(h-(\lambda+2r)^{2}(m+1))c_{r}+\\
+(2r+2)(2\lambda+2r+1)c_{r+1}=0
\end{gathered}
 & r\geq1\\
(h-\lambda^{2}(1+m))c_{0}+2(2\lambda+1)c_{1}=0\,.
\end{cases}\label{eq:recforheun-2}
\end{equation}

We need to determine the convergence of a general power series (\ref{eq:powseries-1})
given by a three-term recursion (\ref{eq:genrecursion-1}). This can
be done by using the simplest form of Poincaré-Perron theory \cite{ronveaux1995heun}.
In a three-term recursion, if, after dividing by suitable function
of $r$, $P_{r}\rightarrow P$, $S_{r}\rightarrow S$ and $R_{r}\rightarrow R$
as $r\rightarrow\infty$, we can define the following characteristic
equation:
\begin{equation}
P+S\rho+R\rho^{2}=0\,.\label{eq:chareq-1}
\end{equation}
The Poincaré-Perron theory states \cite{ronveaux1995heun} that if
the characteristic equation has two roots $\rho_{1}$ and $\rho_{2}$
with different moduli and if $|\rho_{1}|<|\rho_{2}|$ then generically
the series (\ref{eq:powseries-1}) has a radius of convergence of
$|\rho_{2}|^{-1}$. However, if $S_{r}$ can be written as $S_{r}=h-Q_{r}$,
where $h$ does not depend on $r$, then for certain values of $h$
the series (\ref{eq:powseries-1}) converges up to $|\rho_{1}|^{-1}$.
These values of $h$ can be given by a continued fraction equation
\cite{ronveaux1995heun}:
\begin{equation}
h-Q_{0}=\frac{R_{0}P_{1}}{h-Q_{1}-\frac{R_{1}P_{2}}{h-Q_{2}-\dots}}\,.\label{eq:contfrac-1}
\end{equation}
In our case, the characteristic equation (\ref{eq:chareq-1}), $m-(m+1)\rho+\rho^{2}=0$,
has the roots $\rho_{1}=1$ and $\rho_{2}=m$. So, unless $h$ is
given by the continued fraction (\ref{eq:contfrac-1}), the series
(\ref{eq:powseries-1}) has a convergence radius of $1$. However,
it can be extended to $\frac{1}{m}$ if $h$ satisfies (\ref{eq:contfrac-1}),
which is the quantisation of momenta.

Let us now extend the analysis for the odd solutions for our equation
(\ref{eq:sninverse-1}), which is searched for in the form of
\begin{equation}
\psi=\mathrm{sgn}(z)\mathrm{cn}(z,m)|\mathrm{sn}(z,m)|^{\lambda}\tilde{\psi}\,.
\end{equation}
After performing the same computations as in the even case, we get
the the following solution:
\begin{equation}
\begin{gathered}\tilde{\psi}(z)=Hl\left(\frac{1}{m},\frac{\lambda^{2}+\frac{(\lambda+1)^{2}}{m}-\frac{h}{m}}{4},\frac{\lambda+1}{2},\frac{\lambda+2}{2},\frac{1+2\lambda}{2},\frac{3}{2},\text{sn}^{2}(z)\right)\end{gathered}
\label{eq:heunoddsmall-1}
\end{equation}
If we consider again the solutions as infinite power series we get
the following recursion:
\begin{equation}
\begin{cases}
\begin{gathered}m(\lambda+2r-1)(\lambda+2r)c_{r-1}+(h-(\lambda+2r+1)^{2}-m(\lambda+2r)^{2})c_{r}+\\
+(2r+2)(2\lambda+2r+1)c_{r+1}=0
\end{gathered}
 & r\geq1\\
(h-(\lambda+1)^{2}-m\lambda{}^{2}))c_{0}+2(2\lambda+1)c_{1}=0\,.
\end{cases}\label{eq:recforheun-1-1}
\end{equation}
The quantization of the momentum comes from the same continued fraction
equation (\ref{eq:contfrac-1}).

\subsection{Correspondence between the eigenvalues of the elliptic potential
and the eigenvalues of the Lamé equation}

Interestingly, the admitted values $h$ for integer $\lambda$ correspond
to certain eigenvalues of the Lamé equation. To see this, let us write
the Lamé equation's Jacobian form \cite{ince1940v}:
\begin{equation}
-\frac{\partial^{2}y(\tilde{z})}{\partial\tilde{z}^{2}}+\lambda(\lambda-1)m\:\text{sn}^{2}(\tilde{z})y(\tilde{z})=hy(\tilde{z})\,.
\end{equation}
Note that the literature uses the notation $n=\lambda-1$. 

Let us start with the even solutions (\ref{eq:heunevensmall-1}).
In order to see the correspondence we use a power series expansion
of the form \cite{ince1940v}:
\begin{equation}
y(\tilde{z})=\mathrm{cn}\tilde{z}\:\mathrm{dn}\tilde{z}\sum_{r=0}^{\infty}c_{r}\mathrm{sn}^{2r+1}\tilde{z}\,.
\end{equation}
 and also demand convergence upto ${\rm sn}(\tilde{z})\sim1$. Plugging
back the solution into the differential equation we get the following
recursion \cite{ince1940v}:
\begin{equation}
m(2r-\lambda+2)(\lambda+2r+1)c_{r-1}+(h-(2r+2)^{2}(m+1))c_{r}+(2r+2)(2r+3)c_{r+1}=0\,.
\end{equation}
For even values of $\lambda$ we can substitute $2r=\lambda-2+2r'$
to get
\begin{equation}
2r'm(2\lambda+2r'-1)c_{r-1}+(h-(\lambda+2r')^{2}(m+1))c_{r}+(\lambda+2r')(\lambda+2r'+1)c_{r+1}=0\,.\label{eq:recoddlame1-1}
\end{equation}
which gives the same continued fraction as eq. (\ref{eq:recforheun-2}).
Having the same continued fraction expansion they lead to the same
requirements for the values $h$ as in the Heun case. Note that these
eigenfunctions are denoted as $b_{\lambda-1}^{2s+2}$, while the eigenvalues
of the odd Lamé functions as $Es_{\lambda-1}^{2s+2}(z)$ \cite{ince1940v}.
Here $s$ labels the solutions. 

We can get the correspondence for odd $\lambda$ values by looking
for Lamé functions in the form of \cite{ince1940v}
\begin{equation}
y(\tilde{z})=\mathrm{cn}\tilde{z}\mathrm{\:dn}\tilde{z}\sum_{r=0}^{\infty}c_{r}\mathrm{sn}^{2r}\tilde{z}\,.
\end{equation}
The recursion is given by
\begin{equation}
m(2r-\lambda+1)(\lambda+2r)c_{r-1}+(h-(2r+1)^{2}(m+1))c_{r}+(2r+1)(2r+2)c_{r+1}=0\,.
\end{equation}
which by setting $2r=\lambda-1+2r'$ can be transformed to 
\begin{equation}
2r'm(2\lambda+2r'-1)c_{r-1}+(h-(\lambda+2r')^{2}(m+1))c_{r}+(\lambda+2r')(\lambda+2r'+1)c_{r+1}=0\,,
\end{equation}
which is the same recursion as eq. (\ref{eq:recoddlame1-1}). These
eigenvalues are denoted as $a_{\lambda-1}^{2s+1}$ and correspond
to the even Lamé functions $Ec_{\lambda-1}^{2s+1}(z)$ \cite{ince1940v}. 

Now, we should turn our attention to the odd solutions (\ref{eq:heunoddsmall-1}).
We can show the correspondence with the Lamé eigenvalues for integer
$\lambda$ values again. First, to show for even $\lambda$'s we write
the solution of the Lamé functions in form of \cite{ince1940v}
\begin{equation}
y(\tilde{z})=\mathrm{cn}\tilde{z}\sum_{r=0}^{\infty}c_{r}\mathrm{sn}^{2r}\tilde{z}\,.
\end{equation}
We get the following recursion \cite{ince1940v}:
\begin{equation}
m(2r-\lambda)(\lambda+2r+1)c_{r-1}+(h-(2r+1)^{2}-4r^{2}m)c_{r}+(2r+1)(2r+2)c_{r+1}=0\,.
\end{equation}
Setting $2r=\lambda+2r'$ we get
\begin{equation}
\begin{gathered}2r'm(2\lambda+2r'-1)c_{r-1}+(h-(\lambda+2r'+1)^{2}-(\lambda+2r')^{2}m)c_{r}+\\
+(\lambda+2r'+1)(\lambda+2r'+2)c_{r+1}=0\,,
\end{gathered}
\label{eq:lamereceven2-1}
\end{equation}
which confirms the correspondence again. This time, the involved Lamé
eigenvalues are $a_{\lambda-1}^{2s+1}$ and correspond to the even
Lamé functions $Ec_{\lambda-1}^{2s+1}(z)$ \cite{ince1940v}.

To show for odd $\lambda$'s we need the following type of Lamé solution
\cite{ince1940v}:
\begin{equation}
y(z)=\mathrm{cn}z\sum_{r=0}^{\infty}c_{r}\mathrm{sn}^{2r+1}z\,.
\end{equation}
The recursion following from this form is \cite{ince1940v}
\begin{equation}
m(2r-\lambda+1)(\lambda+2r)c_{r-1}+(h-(2r+2)^{2}-(2r+1)^{2}m)c_{r}+(2r+2)(2r+3)c_{r+1}=0\,.
\end{equation}
Again, let us substitute $2r=\lambda-1+2r'$:
\begin{equation}
\begin{gathered}2r'm(2\lambda+2r'-1)c_{r-1}+(h-(\lambda+2r'+1)^{2}-(\lambda+2r')^{2}m)c_{r}+\\
+(\lambda+2r'+1)(\lambda+2r'+2)c_{r+1}=0\,,
\end{gathered}
\end{equation}
which is exactly the same recursion as eq. (\ref{eq:lamereceven2-1}).
The corresponding eigenvalues are $b_{\lambda-1}^{2s+2}$ corresponding
to the odd Lamé functions $Es_{\lambda-1}^{2s+2}(z)$ \cite{ince1940v}.

\bibliographystyle{elsarticle-num}
\bibliography{RSsG}

\end{document}